\documentclass{article}

\usepackage{arxiv}

\usepackage{amssymb}
\usepackage{amsmath}
\usepackage[utf8]{inputenc} % allow utf-8 input
\usepackage[T1]{fontenc}    % use 8-bit T1 fonts
\usepackage{hyperref}       % hyperlinks
\usepackage{url}            % simple URL typesetting
\usepackage{booktabs}       % professional-quality tables
\usepackage{amsfonts} 
\usepackage{subcaption}
\usepackage{nicefrac}       % compact symbols for 1/2, etc.
\usepackage{microtype}      % microtypography
\usepackage{bbm, bm}
\newcommand{\bs}[1]{\boldsymbol{#1}}
\usepackage{color}
\usepackage{graphbox}
\graphicspath{ {./images/} }
\usepackage{lineno}
\usepackage{mathtools}
\usepackage{siunitx}

\usepackage{marvosym}
\usepackage{microtype}      % microtypography
\usepackage{multirow}
\usepackage{nicefrac}       % compact symbols for 1/2, etc.

\usepackage{tabularx}
\usepackage[title,titletoc,toc]{appendix}
\usepackage{upgreek}
\usepackage{url}         % simple URL typesetting

\usepackage{xcolor}

\newcommand\mika[1]{{\color{black}{{#1}}}}
\title{Modelling multivariate spatio-temporal data with identifiable variational autoencoders}

\author{Mika Sipilä \\
	Department of Mathematics and Statistics\\
	University of Jyvaskyla\\
	Finland \\
	\And
        Claudia Cappello \\
        DSE - Section of Mathematics and Statistics \\
        University of Salento \\
        Italy \\
        \And
        Sandra De Iaco \\
        DSE - Section of Mathematics and Statistics \\
        University of Salento \\
        Italy \\
        \And
	Klaus Nordhausen \\
	Department of Mathematics and Statistics\\
	University of Jyvaskyla\\
	Finland \\
        \And
        Sara Taskinen \\
	Department of Mathematics and Statistics\\
	University of Jyvaskyla\\
	Finland \\
}

\begin{document}

\maketitle

\begin{abstract}
Modelling multivariate spatio-temporal data with complex dependency structures is a challenging task but can be simplified by assuming that the original variables are generated from independent latent components. If these components are found, they can be modelled univariately. Blind source separation aims to recover the latent components by estimating the unknown linear or nonlinear unmixing transformation based on the observed data only. In this paper, we extend recently introduced identifiable variational autoencoder to the nonlinear nonstationary spatio-temporal blind source separation setting and demonstrate its performance using comprehensive simulation studies. Additionally, we introduce two alternative methods for the latent dimension estimation, which is a crucial task in order to obtain the correct latent representation. Finally, we illustrate the proposed methods using a meteorological application, where we estimate the latent dimension and the latent components, interpret the components, and show how nonstationarity can be accounted and prediction accuracy can be improved by using the proposed nonlinear blind source separation method as a preprocessing method.
\end{abstract}


\section{Introduction}
\label{sec:intro}

Many real world phenomena, such as weather, epidemiological patterns and ecosystem dynamics, are multivariate spatio-temporal, meaning that multivariate observations $\bs x(\bs s, t) \coloneqq \bs x \in \mathbb{R}^S$ are observed in a spatial location $\bs s \in \mathcal{S} \subset \mathbb{R}^D$ at time $t \in \mathcal{T} \subset{\mathbb{R}}$, where $\mathcal{S}$ is called a spatial domain, $\mathcal{T}$ is called a temporal domain and $D$ is a spatial dimension. Without loss of generality, we assume from now on that $D=2$. A multivariate observation $\bs x$ contains measurements of multiple, usually dependent, random variables describing the phenomenon of interest. When modelling such multivariate spatio-temporal data, one has to account not only the dependence between the variables, but also the dependences in space and in time. The dependence structure is often described through spatio-temporal covariance function $\bs C(\bs x(\bs s, t), \bs x(\bs s', t'))$, where $(\bs s, t)$ and $(\bs s', t')$ are two spatio-temporal locations. The covariance $\bs C$ is a $S \times S$ matrix valued functional with elements $C_{ij}$, $i,j=1,\dots,S$, defined as $C_{ij} = C(x_i(\bs s, t), x_j (\bs s', t')) = E(x_i(\bs s, t)x_j (\bs s', t')) - E(x_i(\bs s, t))E(x_j(\bs s', t'))$. Modelling the covariance function $\bs C$ is usually a highly demanding task, and often, in order to make the modelling feasible, some severely restricting assumptions, such as stationarity or separability, are made. When the spatio-temporal field is assumed to be stationary, the covariance function can be simplified to 
\begin{align}
    \bs C(\bs x(\bs s, t), \bs x(\bs s', t')) = \bs C(\|\bs s - \bs s'\|, |t - t'|),
    \label{eq:stationary_covariance}
\end{align} 
meaning that the value of the function depends only on the distance between the spatial locations and the distance between temporal locations. \mika{If (\ref{eq:stationary_covariance}) does not hold, the data are nonstationary, meaning that the covariance function may differ when spatial or temporal location is altered.} When separability is assumed, the spatio-temporal covariance function can be written as a product of spatial and temporal covariance functions as
\begin{align*}
    \bs C(\bs x(\bs s, t), \bs x(\bs s', t')) = \bs C_\mathcal{S}(\bs s, \bs s')\bs C_\mathcal{T}(t, t'),
\end{align*}
meaning that the spatial and temporal covariance models can be fitted independently and that the spatio-temporal interaction is not considered. The assumptions of stationarity or separability often lead to unrealistically simple models that hence produce nonoptimal results \mika{under nonstationary or nonseparable data}. Accounting complex nonseparable and nonstationary correlation structures is complicated already in the univariate case, for which an overview can be found in \cite{chen2021space}. For multivariate data, the task is even more demanding and computationally challenging as the cross-dependencies between the variables have to be taken into account. For more details of complexity of nonstationary covariance functions for multivariate spatio-temporal data, see \cite{salvana2020nonstationary,PorcuFurrerNychka2021}.

Another approach to simplify the modelling is to assume that the observations are composed of $P$ latent, mutually independent components (ICs) $\bs z(\bs s, t) \coloneqq \bs z \in \mathbb{R}^P$ through some mixing environment. \mika{The main motivation for assuming the ICs is, that} if the latent components $\bs z$ are recovered, they can be modelled univariately making for example nonstationarity much easier to account for. \mika{Being able to model components univariately is especially desirable in spatio-temporal settings, where multivariate modelling is highly demanding and computationally challenging as discussed previously. Additionally, the ICs may reveal some meaningful patterns and structures in the observed data that can lead to new insights of the phenomenon of interest.} A linear blind source separation (BSS) \cite{comon2010handbook} is a popular approach to recover the latent components $\bs z$. In linear BSS, it is assumed that the mixing environment is linear and usually also that $S=P$ meaning that a $P$-variate observable random vector $\bs x=(x_1,\dots,x_P)^\top$ is generated as 
\begin{align}
    \bs x = \bs A \bs z,
\label{eq:linear_bss}
\end{align}
where $\bs A$ is an invertible $P \times P$ mixing matrix and $\bs z = (z_1, \dots, z_P)^\top$ are the $P$-variate latent components. The objective is to recover $\bs A$ and $\bs z$ using only $\bs x$ and varying assumptions on $\bs z$ depending on the method used. For example, spatial BSS (SBSS) \cite{NordhausenOjaFilzmoserReimann2015, bachoc2018spatial}, which is a method for multivariate stationary spatial data, assumes spatially stationary $\bs z$, and a nonstationary extension of SBSS, spatial nonstationary source separation (SNSS) \cite{MuehlmannBachocNordhausen2021}, assumes $\bs z$ to have nonstationary spatial covariance function. SBSS and SNSS recover the latent components by jointly diagonalizing two or more moment-based matrices. Recently, SBSS was also extended for stationary spatio-temporal data yielding spatio-temporal BSS (STBSS) \cite{MuehlmannDeIacoNordhausen2022}. A drawback of STBSS and linear BSS methods in general is that they assume linear mixing (\ref{eq:linear_bss}) which may be too restrictive assumption for many real life applications. %Linear BSS methods usually also assume that the observed data $\bs x$ and the latent components $\bs z$ are both $P$-dimensional and require additional dimension reduction preprocessing step if the observed dimension $S$ is greater than the latent dimension $P$. 
Similarly, the assumption that there are as many latent ``signal'' components as observed variables is in many applications undesirable and it is often hoped that there are significantly fewer signals. This assumption is often needed simply due to the lack of tools for estimating the correct number of signals.
Finally, STBSS is developed only for stationary data, and to our knowledge, there are no spatio-temporal alternatives available for nonstationary data cases.

Recent advancements in unsupervised deep learning, such as variational autoencoders (VAEs) \cite{kingma2013auto} \mika{and generative adverserial networks (GANs) \cite{gan}}, have increased interest for developing nonlinear BSS methods, where the mixing function is not restricted to be linear, but can be any injective function $\bs f:\mathbb{R}^P \rightarrow \mathbb{R}^S$, which generates the observed data $\bs x$ as
\begin{align}
    \bs x = \bs f(\bs z).
\label{eq:nonlinearICA}
\end{align}
The objective is then to identify an unmixing transformation $\bs q:\mathbb{R}^S \rightarrow \mathbb{R}^P$, which returns the latent components $\bs z$ as
\begin{align*}
    \bs z = \bs q(\bs x)
\end{align*}
based on the observations $\bs x$ only. Without any additional assumptions on the mixing transformations $\bs f$ or on the latent components $\bs z$, the model is unidentifiable as there exists infinite nonlinear transformations to generate mutually independent components from the observations \cite{hyvarinen1999nonlinear}. For this reason both VAEs \mika{and GANs}, in general, suffer from the unidentifiability issue. However, in many recent studies \cite{HyvarinenMorioka2016, HyvarinenMorioka2017, Hyvarinen2019, HalvaHyvarinen2020, Khemakhem2020} the identifiability have been achieved by introducing some constraints on the distribution of the latent components $\bs z$. The main assumption leading to identifiability is that the components $z_1,\dots, z_P$ are statistically dependent on a $m$-dimensional auxiliary variable $\bs u$, and that the components are conditionally independent yielding the joint distribution $p(\bs z|\bs u) = \prod_{i=1}^P p(z_i| \bs u)$. In previous studies, the main focus has been in time series data for which several \mika{algorithms and} examples for auxiliary variables exist in the literature. \mika{In case of stationary time series data, permutation contrastive learning (PCL) \cite{HyvarinenMorioka2017} can be used, for which $\bs u$ is usually given by one or more previous observations in time. For nonstationary time series data, the available methods are time contrastive learning (TCL) \cite{HyvarinenMorioka2016}, hidden Markov nonlinear ICA (HM-NICA) \cite{HalvaHyvarinen2020} and temporal identifiable VAE (iVAE) \cite{Khemakhem2020}, all of which use the time segment of the observation as $\bs u$. Generalized contrastive learning \cite{Hyvarinen2019} and nonlinear ICA with switching linear dynamical systems ($\Delta$-SNICA) \cite{halva2021disentangling} can account both stationary and nonstationary time series. In HM-NICA and $\Delta$-SNICA, the auxiliary variables $\bs u$ are not explicitly provided by the user, but they are instead assumed to be hidden states that are modelled simultaneously by the algorithms. In \cite{SipilaNonlinearSBSS}, iVAE was studied further and extended to nonstationary spatial setting, where spatial segmentation was used as $\bs u$. \cite{halva2021disentangling} also introduced a structured nonlinear ICA framework which could be used for spatial process, but did not provide any algorithm for the method. In addition to these more general identifiable nonlinear BSS methods, many other deep learning based BSS methods \cite{ANSARI2023126895} have been introduced for mainly acoustic signal specific settings, where only serial dependence is present. However, the spatio-temporal data as discussed in this paper is special in sense that in the temporal domain there is natural direction of dependence (past-future) while in the spatial domain such direction is missing and the dependence is usually considered as a function of the distance between two points. Hence, none of the previous methods are directly applicable or optimal for such spatio-temporal data. Note that regularly spaced spatio-temporal data is often represented as tensor data, and BSS methods developed for such specific cases, like those in \cite{VirtaNordhausen2017}, are generally not applicable to broader spatio-temporal settings.
} 

\mika{In particular, we are interested in iVAE, which} utilizes the auxiliary variable to make VAE identifiable. iVAE is capable of estimating nonlinear injective mixing function, meaning that it allows the latent dimension $P$ to be less or equal to the observed dimension $S$. However, the latent dimension $P$ has to be estimated beforehand, and currently the nonlinear BSS framework lacks methods for the latent dimension estimation. 

In this paper, iVAE is extended to nonstationary spatio-temporal setting by introducing three novel approaches to construct the auxiliary variables. \mika{
The proposed methods address two key limitations of previous STBSS approaches: they accommodate nonlinear mixing functions and allow for more observed variables than latent components. Moreover, the developed methods are suitable for nonstationary data, unlike earlier STBSS methods, which rely on the assumption of stationarity.
} The three developed methods, coordinate based, segmentation based and radial basis function based iVAE algorithms, are studied using comprehensive simulation studies to find how various types of nonstationarity affect the performance of the methods. The best performing method, radial basis function based iVAE, is illustrated in real life meteorological application where the recovered latent components are interpreted, and a new procedure to account for nonstationarity in modelling and predicting multivariate data is demonstrated. Moreover, nonlinear BSS framework currently lacks methods for estimating the latent dimension $P$, which is a crucial task in order to recover the true latent components and to obtain as low dimensional representation of the data as possible without losing much information. Therefore, two alternative procedures for latent dimension estimation are introduced. \mika{To conclude, the main contributions of this paper are: 
\begin{enumerate} \item Extending iVAE to the nonstationary spatio-temporal setting by proposing three novel approaches for constructing auxiliary variables. \item Introducing two alternative procedures for latent dimension estimation. \item Developing a new iVAE-based method for addressing nonstationarity in the modeling and prediction of spatio-temporal data. \end{enumerate}}
The rest of the paper is organized as follows. In Section~\ref{sec:ivae} we review basic theory behind VAE and iVAE, and discuss the identifiability, after which the spatio-temporal iVAE extensions are introduced in Section~\ref{sec:ivae_stbss}. In Section~\ref{sec:simulations}, the introduced methods are compared using simulation studies, and two alternative latent dimension estimation methods are studied. Finally, Section~\ref{sec:case_study} shows a real data example and Section~\ref{sec:conclusion} concludes the paper.   
 
\section{Variational autoencoders and identifiability}
\label{sec:ivae}

Let $\bs x \in \mathbb{R}^S$ be an observable random vector and $\bs z \in \mathbb{R}^P$, $P \leq S$, be a latent random vector, i.e., a source vector. Variational autoencoders (VAE) \cite{kingma2013auto} assume that the observed data are generated from a deep latent variable model with the structure %of 
\begin{align*}
    p^*(\bs x, \bs z) = p^*(\bs x|\bs z) p^*(\bs z),
\end{align*}
where $p^*$ is a true, unknown generative distribution, $\bs z \sim p^*(\bs z)$ and $\bs x \sim p^*(\bs x | \bs z)$. The distribution of the observed data is then obtained as
\begin{align*}
    p^*(\bs x) = \int p^*(\bs x, \bs z) d\bs z.    
\end{align*}
VAE  %is 
consists of an encoder $\bs g(\bs x)$ and a decoder $\bs h(\bs z)$, which are parameterized by deep neural networks with parameters $\bs \theta=(\bs \theta_{\bs g}^\top, \bs \theta_{\bs h}^\top)^\top$. The encoder maps the observed data to mean vector $\bs \mu_{\bs z| \bs x} \in \mathbb{R}^P$ and variance vector $\bs \sigma_{\bs z| \bs x} \in \mathbb{R}^P$, which are used to sample a new latent representation $\bs z'$ by applying the reparametrization trick \cite{kingma2013auto}. The decoder transforms the latent representation $\bs z'$ back to the observable data $\bs x'$. The VAE framework allows effective optimization of the parameters $\bs \theta$ so that after optimization we have that
\begin{align*}
    p_{\bs \theta}(\bs x) \approx p^*(\bs x).
\end{align*}

\begin{figure}
  \centering
  \includegraphics[width=0.9\textwidth]{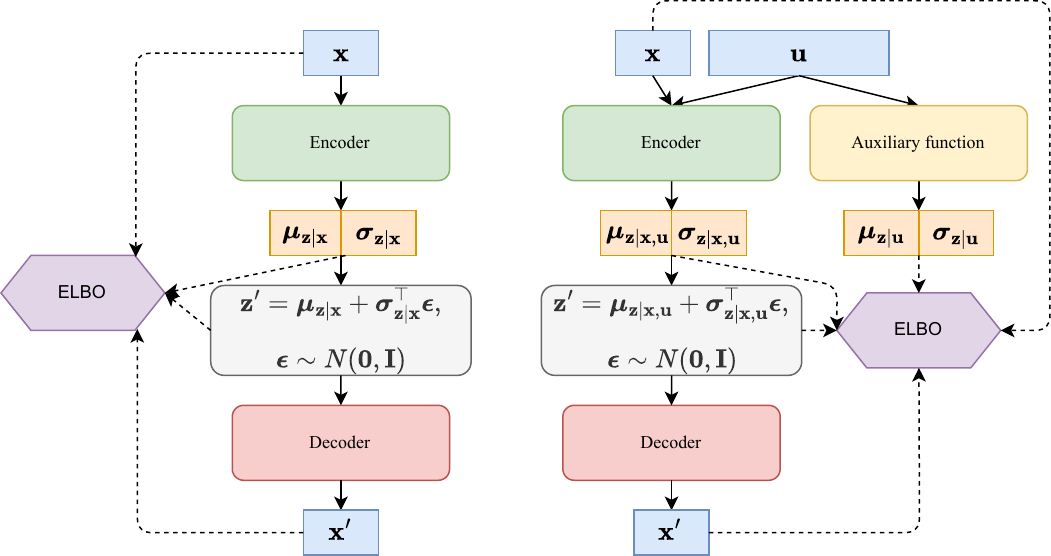}
  \caption{Schematic representations of VAE (left) and iVAE (right) models. For VAE, the lower bound of the data log likelihood (ELBO) is formed of $\bs x$, $\bs x'$, $\bs z'$, $\bs \mu_{\bs z|\bs x}$ and $\bs \sigma_{\bs z|\bs x}$. In iVAE, ELBO has in addition $\bs \mu_{\bs z|\bs u}$ and $\bs \sigma_{\bs z|\bs u}$ which are provided by the auxiliary function. The latent components are obtained as $\bs \mu_{\bs z | \bs x, \bs u}$.}
  \label{fig:vaeivae}
\end{figure}

The VAE framework learns the full generative model $p_{\bs \theta}(\bs x, \bs z) = p_{\bs \theta}(\bs x|\bs z) p_{\bs \theta}(\bs z)$ and a variational approximation $q_{\bs \theta}(\bs z | \bs x)$ of the posterior distribution $p_{\bs \theta}(\bs z | \bs x)$ by maximizing the lower bound of the data log-likelihood, or evidence lower bound (ELBO), defined as
\begin{align*}
    \mathcal{L}(\bs \theta | \bs x) \geq 
    E_{q_{\bs \theta}(\bs z|\bs x)}\big (\text{log}\, p_{\bs \theta}(\bs x | \bs z)  + \text{log}\,p_{\bs \theta}(\bs z) - \text{log}\,q_{\bs \theta}(\bs z | \bs x) \big)
\end{align*}
with respect to the parameter vector $\bs \theta$. The problem however is that the model is not identifiable, meaning that even though we have a good estimate of the marginal distribution $p^*(\bs x)$, there is no guarantee that $p_{\bs \theta}(\bs x, \bs z) \approx p^*(\bs x, \bs z)$. More formally, the model is identifiable if for all $(\bs x, \bs z)$ it holds that
\begin{align*}
    \forall (\bs \theta, \bs \theta'): p_\theta(\bs x) = p_{\theta'}(\bs x) \implies  p_{\bs \theta}(\bs x, \bs z) = p_{\bs \theta'}(\bs x, \bs z).
\end{align*}
This means that if we find a parameter vector $\bs \theta$ for which $p_\theta(\bs x) = p^*(\bs x)$, we also have that $p_{\bs \theta}(\bs x, \bs z) = p^*(\bs x, \bs z)$. This leads to the fact that we have found the correct source density distribution $p_{\bs \theta}(\bs z) = p^*(\bs z)$ and correct conditional distributions $p_{\bs \theta}(\bs x | \bs z) = p^*(\bs x | \bs z)$ and $p_{\bs \theta}(\bs z | \bs x) = p^*(\bs z | \bs x)$. The whole VAE model is illustrated in Figure~\ref{fig:vaeivae}. 
%If the identifiability applied for VAE, we would get the variational approximation $q_{\bs \theta}(\bs z | \bs x)$ of the true distribution $p^*(\bs z | \bs x)$.

In the nonlinear BSS framework, the identifiability has been recently achieved by assuming that the latent sources $\bs z$ have a conditional distribution $p(\bs z | \bs u)$, where $\bs u \in \mathbb{R}^m$ is an auxiliary variable. The auxiliary variable can for example be previous observations in time \cite{HyvarinenMorioka2017} or current time index \cite{HyvarinenMorioka2016, HalvaHyvarinen2020, Hyvarinen2019}. Similarly, by assuming that the true latent generating model has the form %of
\begin{align}
\label{eq:iVAE_generative_model}
    p^*(\bs x, \bs z | \bs u) = p^*(\bs x|\bs z) p^*(\bs z | \bs u),
\end{align}
the identifiability can be achieved in the VAE framework, yielding identifiable VAE (iVAE) \cite{Khemakhem2020}. In iVAE, the distribution $p^*(\bs x|\bs z)$ is defined as 
\begin{align*}
    p^*(\bs x|\bs z) = p_{ \bs \epsilon}^*(\bs x - \bs f(\bs z)),
\end{align*}
which means that $\bs x$ can be decomposed into $\bs x = \bs f(\bs z) + \bs \epsilon$, where $\bs \epsilon$ is an independent noise vector with density $p_{\bs \epsilon}$. Assuming the nonnoisy nonlinear BSS model (\ref{eq:nonlinearICA}), the distribution $p_{\bs \epsilon}$ can be modelled with Gaussian distribution with infinitesimal variance. The function $\bs f: \mathbb{R}^P \rightarrow \mathbb{R}^S$ is an injective, but possibly nonlinear function. The conditional distribution of latent sources $\bs z$ is assumed to be a part of the exponential family, that is, 
\begin{align}
    p_{\bs T, \bs \lambda}(\bs z| \bs u) = \prod_{i=1}^P \frac{Q_i(z_i)}{Z_i(\bs u)} \text{exp}\left[ \sum_{j=1}^k T_{i,j}(z_i) \lambda_{i,j}(\bs u) \right],
    \label{eq:sourcedist}
\end{align}
where $Q_i(z_i)$ is a base measure, $Z_i(\bs u)$ is a normalizing constant, 
$\bs T_i(z_i)=(T_{i,1}(z_i), \dots, T_{i,k}(z_i))^\top$ contains sufficient statistics, and $\bs \lambda_i(\bs u)=(\lambda_{i,1}(\bs u), \dots, \linebreak \lambda_{i,k}(\bs u))^\top$ contains the parameters depending on $\bs u$. The dimension $k$ of each sufficient statistic $\bs T_i(z_i)$ and $\bs \lambda_i(\bs u)$ is assumed to be fixed. The latent components $\bs z$ are identifiable up to permutation and signed scaling under some generally mild conditions on the mixing function $\bs f$, the sufficient statistics $\bs T$ and the auxiliary variable $\bs u$. In this study, we construct iVAE assuming Gaussian latent components. Then, for identifiability, the variances of the latent components are required to vary enough based on the auxiliary variable $\bs u$ and the mixing function $\bs f$ is required to have continuous partial derivatives. The exact identifiability conditions can be found in \cite{Khemakhem2020}.

The iVAE model is similar to the regular VAE model with the exception that iVAE has an additional auxiliary function $\bs w(\bs u)$ and its parameters $\bs \theta_{\bs w}$ to be estimated, and the encoder $\bs g(\bs x, \bs u)$ takes both, observations $\bs x$ and the auxiliary variables $\bs u$ as an input. The auxiliary function $\bs w$ maps $\bs u$ into $\bs \mu_{\bs z| \bs u}$ and $\bs \sigma_{\bs z| \bs u}$, which are used to calculate the loss. For iVAE model, ELBO is obtained as
\begin{align*}
    \mathcal{L}(\bs \theta | \bs x, \bs u) &\geq 
    E_{q_{\bs \theta}(\bs z|\bs x, \bs u)}\big (\text{log}\, p_{\bs \theta_{\bs h}}(\bs x | \bs z) + \text{log}\,p_{\bs \theta_{\bs w}}(\bs z | \bs u) - \text{log}\,q_{\bs \theta_{\bs g}}(\bs z | \bs x, \bs u) \big),
\end{align*}
where $\text{log}\, p_{\bs \theta_{\bs h}}(\bs x | \bs z)$ controls the reconstruction accuracy and $\text{log}\,p_{\bs \theta_{\bs w}}(\bs z | \bs u) - \text{log}\,q_{\bs \theta_{\bs g}}(\bs z | \bs x, \bs u)$ is a Kullback-Leibler (KL) divergence between $p_{\bs \theta_{\bs w}}(\bs z | \bs u)$ and $q_{\bs \theta_{\bs g}}$ keeping the distributions $p_{\bs \theta_{\bs w}}$ and $q_{\bs \theta_{\bs g}}$ as similar as possible. ELBO is maximized to obtain the estimated parameters $\bs \theta=(\bs \theta_{\bs g}^\top, \bs \theta_{\bs h}^\top, \bs \theta_{\bs w}^\top)^\top$. 
The distributions $p_{\bs \theta_{\bs h}}$, $p_{\bs \theta_{\bs w}}$ and $q_{\bs \theta_{\bs g}}$ are typically Gaussian distributions, where the functions $\bs h$, $\bs w$ and $\bs g$ give the mean and variance parameters of the distributions. The distributions can also be other than Gaussian as long as the resampling can be done using the reparametrization trick to allow the backpropagation go through the resampling node. Then, the functions $\bs h$, $\bs w$ and $\bs g$ do not give mean and variance, but the parameters according the chosen distributions. As we assume Gaussian latent data in this paper, we have $p_{\bs \theta_{\bs w}} = N(\bs z| \bs \mu_{\bs z|\bs u}, \text{diag}(\bs \sigma_{\bs z|\bs u}))$, $q_{\bs \theta_{\bs g}} = N(\bs z| \bs \mu_{\bs z|\bs x, \bs u}, \text{diag}(\bs \sigma_{\bs z|\bs x, \bs u}))$ and $p_{\bs \theta_{\bs h}} = N(\bs x| \bs x', \beta \bs I)$, where $\beta > 0$ is a small constant as $p_{\bs\theta_{\bs h}}(\bs x | \bs z)$ estimates the true distribution $p^*(\bs x | \bs z)$ with infinitesimal variance. By increasing $\beta$, the weight of the reconstruction accuracy in the ELBO decreases. Based on our empirical investigations, we use $\beta = 0.02$ which provides a good balance between the reconstruction error and KL divergence in ELBO, and leads to good performance. Figure~\ref{fig:vaeivae} has representations of both VAE and iVAE models and illustrates the difference between the models.

\section{iVAE for STBSS}
\label{sec:ivae_stbss}

To perform nonlinear spatio-temporal blind source separation using iVAE, the auxiliary variables $\bs u$ must be selected appropriately. The main assumption for identifiability in spatio-temporal setting is that the variances of the latent components are varying in space and/or in time. This assumption is met by assuming that the latent components are second-order nonhomogeneous, meaning that the second moment of the marginal distribution $p(z_i)$ is not invariant with respect to the location shift in space and/or in time. In addition, the latent components are allowed to be first-order nonhomogeneous, meaning that the components can have nonconstant spatio-temporal trend. The auxiliary variables are constructed in a way that the auxiliary function $\bs w$ is capable to learn and estimate the mean and the variance vectors of the location of the corresponding multivariate observation. We propose three spatio-temporal iVAE methods; a naive coordinate based method, a segmentation based method extended from spatial iVAE \cite{SipilaNonlinearSBSS} and a radial basis function based method utilizing ideas of \cite{spatiotemporal_deepkriging}. Each of the three methods construct the auxiliary data differently based on the spatio-temporal location of the observation. Notice that although in many options below the auxiliary variables are constructed separately for spatial coordinates and temporal coordinates, the auxiliary functions can still learn complex spatio-temporal interactions in the mean and in the variance as the auxiliary functions are modelled by deep neural networks. Furthermore, even though the methods for constructing the auxiliary data are defined here for spatial dimension $D=2$, the same ideas apply also for higher $D$. 

\mika{The approaches presented here are well scalable in terms of computation time. The computation time grows sublinearly with respect to the sample size $n$ (as fewer training epochs are typically needed with larger datasets) and linearly with respect to the dimensions of observed data, latent data and auxiliary data. However, if the dimension of the auxiliary data and sample sizes are large, the memory usage may grow unless the auxiliary data are formed batch-wise. An in-depth analysis of computational complexity is provided in \ref{sec:computational_complexity}.}

\subsection{Coordinate based algorithm}

In coordinate based iVAE, the preprosessed coordinates are used directly as auxiliary variable. The preprosessed coordinates are obtained by applying min-max normalization to each dimension. The preprosessed coordinates are then
\begin{align*}
\tilde{s}_1 = \frac{s_1 - s^{\text{min}}_1}{s^{\text{max}}_1 - s^{\text{min}}_1}, \tilde{s}_2 = \frac{s_2 - s^{\text{min}}_2}{s^{\text{max}}_2 - s^{\text{min}}_2} \text{ and } \tilde{t} = \frac{t - t^{\text{min}}}{t^{\text{max}} - t^{\text{min}} },
\end{align*} 
where $s^{\text{min}}_1, s^{\text{min}}_2$ and $t^{\text{min}}$ are the minimum coordinates of the locations of the observations, and $s^{\text{max}}_1, s^{\text{max}}_2$ and $t^{\text{max}}$ are the maximum coordinates of the locations of the observations. The algorithm with the preprosessed coordinates, $\bs u(\bs s, t) = (\tilde{s}_1, \tilde{s}_2, \tilde{t})^\top$, as auxiliary variable is denoted by iVAEc. 

\subsection{Segmentation based algorithm}

In segmentation based iVAE, a spatio-temporal segmentation is used as an auxiliary variable. The spatio-temporal segmentation means that the domain $\mathcal{S} \times \mathcal{T}$ is divided into $m$ nonintersecting segments $\mathcal{K}_i \in \mathcal{S} \times \mathcal{T}$ so that $\mathcal{K}_i \cap \mathcal{K}_j = \emptyset$ for all $i \neq j$, $i,j = 1,\dots,m$, and $\cup_{i=1}^m \mathcal{K}_i = \mathcal{S} \times \mathcal{T}$. By using an indicator function $\mathbbm{1}$, the auxiliary variable for the observation $\bs x(\bs s, t)$ can be written as $\bs u(\bs s, t) = (\mathbbm{1}( (\bs s, t) \in \mathcal K_1), \dots, \mathbbm{1}( (\bs s, t) \in \mathcal K_m)))^\top$, where $\mathbbm{1} ((\bs s, t) \in \mathcal K_i) = 1$, if the location $(\bs s, t)$ is within the segment $\mathcal K_i$, and otherwise $\mathbbm{1} ((\bs s, t) \in \mathcal K_i) = 0$. This results into $m$-dimensional standard basis vector, where the value 1 gives the spatio-temporal segment in which the location of the observation belongs.

If the spatio-temporal domain is large and small segments are used, the dimension $m$ of the auxiliary variable becomes very large. To lower the dimension, the spatial and temporal segmentations can be considered separately. This means that the auxiliary data is composed of $m_S$ spatial segments $\mathcal{S}_i \in \mathcal{S}$ and $m_T$ temporal segments $\mathcal{T}_i \in \mathcal{T}$ so that $\mathcal{S}_i \cap \mathcal{S}_j = \emptyset$ for all $i \neq j$, $i,j = 1,\dots,m_S$, $\cup_{i=1}^{m_S} \mathcal{S}_i = \mathcal{S}$, $\mathcal{T}_i \cap \mathcal{T}_j = \emptyset$ for all $i \neq j$, $i,j = 1,\dots,m_T$, and $\cup_{i=1}^{m_T} \mathcal{T}_i = \mathcal{T}$. Then, the auxiliary variable for the observation $\bs x(\bs s, t)$ is $\bs u(\bs s, t) = (\mathbbm{1}( \bs s \in \mathcal S_1), \dots, \mathbbm{1}( \bs s \in \mathcal S_{m_S}), \mathbbm{1}( t \in \mathcal T_1), \dots, \mathbbm{1}( t \in \mathcal T_{m_T}) ))^\top$. The auxiliary variable is $(m_S + m_T)$-dimensional and has two nonzero entries for each observation. The dimension can be reduced even further by considering also the $x$-axis and $y$-axis of the spatial domain separately. \mika{Segmentation based auxiliary variables are illustrated in Figure~\ref{fig:spatial_and_radial_aux_variables}, in which spatial and temporal segmentations are considered separately.} We denote the algorithm with all dimensions segmented separately as iVAEs1, with space and time segmented separately as iVAEs2, and with spatio-temporal segmentation as iVAEs3, respectively. 

\subsection{Radial basis function based algorithm}

In radial basis function based iVAE, the auxiliary variable is defined using radial basis functions (see e.g. \cite{hastie2009elements}). The idea is that with large number of appropriate radial basis functions, the model incorporates much more spatio-temporal information than by using the coordinates only. Similar ideas have been used recently in \cite{chen2020deepkriging, spatiotemporal_deepkriging} to perform deep learning based spatial and spatio-temporal predicting by using the spatial and spatio-temporal locations transformed into radial basis functions as input for deep neural networks. Following \cite{spatiotemporal_deepkriging}, we transform spatial and temporal locations separately into radial basis functions. Let $\{ \bs o^{\mathcal{S}}_i \}$, $i = 1, \dots, K_\mathcal{S}$, where $\bs o^{\mathcal{S}}_i \in \mathcal{S}$, be a set of spatial node points, and let $\{o^{\mathcal{T}}_i \}$, $i = 1, \dots, K_\mathcal{T}$, where $o^{\mathcal{T}}_i \in \mathcal{T}$, be a set of temporal node points. The parameter $\zeta$ is a scale parameter. The spatial and temporal radial basis functions are given as
\begin{align*}
    v^\mathcal{S}(\bs s; \zeta, \bs o^\mathcal{S}_i) = v(\| \bs s - \bs o^\mathcal{S}_i \|/\zeta ) \quad \text{and} \quad 
    v^\mathcal{T}(t; \zeta, o^\mathcal{T}_i) = v(| t - o^\mathcal{T}_i |/\zeta ),
\end{align*}
where $v$ is a kernel function such as the Gaussian kernel $v_G(d)=e^{-d^2}$, or one of the Wendland kernels \cite{wendlandkernel1995} such as 
\mika{
\begin{align*}
    v_W(d)=\begin{cases}
    (1-d)^6 (35d^2 + 18d + 3) / 3, & d \in [0, 1] \\
    0, & \text{otherwise}.
\end{cases}
\end{align*}
}
Following \cite{nychka2015multiresolution}, we use a multi-resolution approach to form the spatial and temporal radial basis functions. Each resolution level is composed of its own number of evenly spaced node points and own scaling parameter. A low level resolution with small number of node points and large value of the scaling parameter aims to capture large-scale spatial or temporal dependencies, while a high level resolution with many node points and small scaling parameter aims to find finer details of the dependence structure.

\begin{figure}
    \centering
    \begin{subfigure}[t]{0.48\textwidth}
        \centering
        \includegraphics[width=\linewidth]{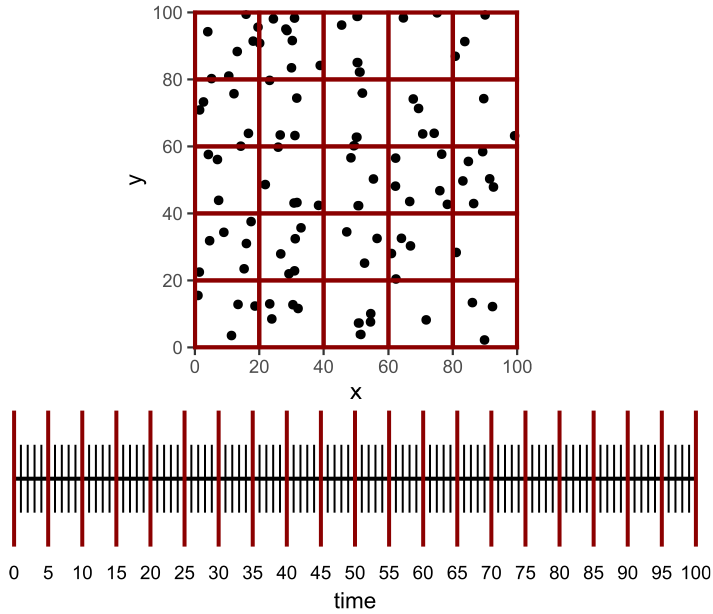}
        \caption{}
    \end{subfigure}\hfill
    \begin{subfigure}[t]{0.48\textwidth}
        \centering
        \includegraphics[width=\linewidth]{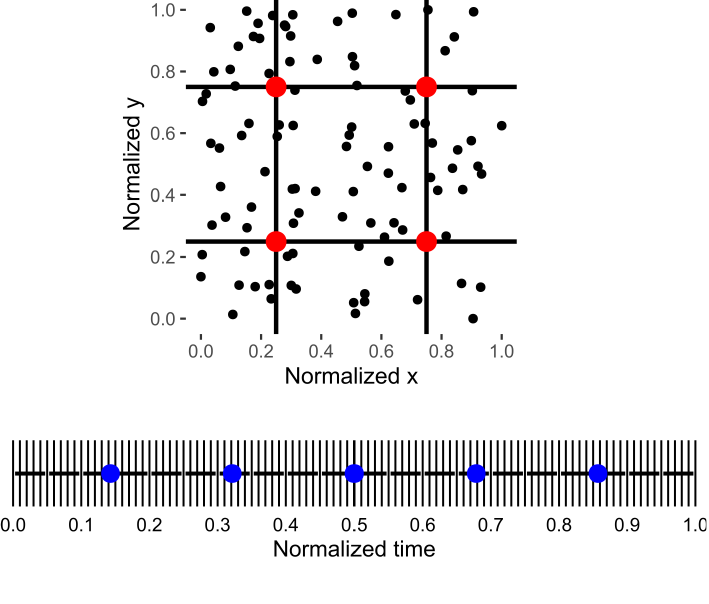}
        \caption{}
    \end{subfigure}
    \begin{subfigure}[t]{0.48\textwidth}
        \centering
        \includegraphics[width=\linewidth]{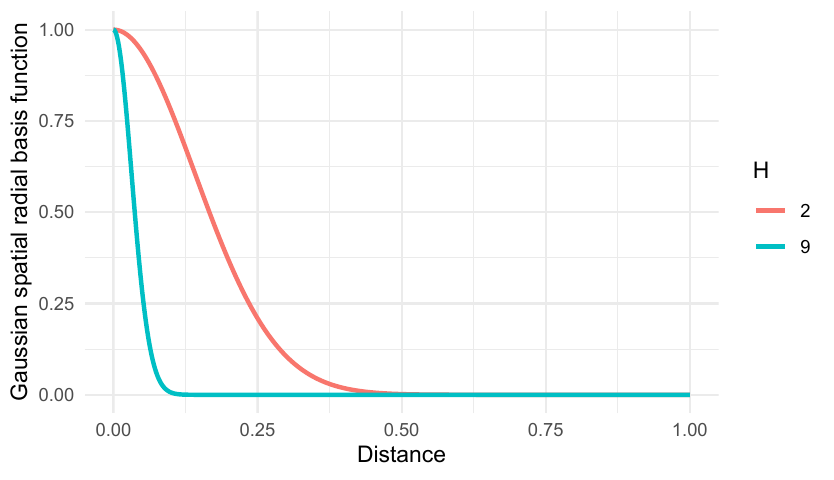}
        \caption{}
    \end{subfigure}
    \begin{subfigure}[t]{0.48\textwidth}
        \centering
        \includegraphics[width=\linewidth]{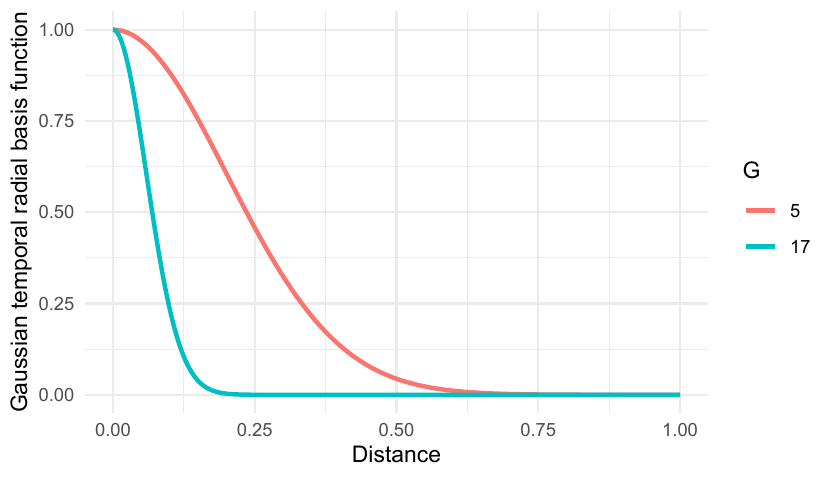}
        \caption{}
    \end{subfigure}
    \caption{\mika{Illustrations of auxiliary variables of segmentation based iVAE (iVAEs2) (a) and radial basis function based iVAE (b). The top figure of (a) illustrates spatial segmentation, where each segment has size $20 \times 20$ producing 25 spatial segments, and the bottom figure illustrates temporal segmentation, where each segment has 5 time points, producing 20 temporal segments. In (b), the black lines in the top figure are the normalized x and y values at $1/(H+2)=1/4$ and $1/(H+2) + 1/H = 3/4$ formed by resolution level $H=2$, and the red points represent the produced spatial node points. The blue points represent temporal node points for resolution level $G=5$. Spatial and temporal Gaussian radial basis functions are illustrated in (c) and (d), respectively. The radial basis functions are functions of distance between node point and a spatial or temporal location.}}
    \label{fig:spatial_and_radial_aux_variables}
\end{figure}

To form the radial basis functions, we first preprocess the spatial \mika{and temporal locations} to range $[0, 1]$ using min-max normalization. A $H$-level spatial resolution is formed of evenly spaced grid of node points $\{ \bs o^\mathcal{S}_i \}$ with spacing $1/H$ and an offset $1/(H + 2)$ before the first node point, meaning that $H$-level resolution has the node points $\{ (i, j) : i,j \in \{\frac{1}{H + 2}, \frac{1}{H + 2} + \frac{1}{H}, \dots, 1 - \frac{1}{H + 2}\} \}$. For example 2-level spatial resolution is then composed of the node points $\{ (i, j) : i,j \in \{0.25, 0.75\} \}$. Similarly, a $G$-level temporal resolution is formed of evenly spaced one dimensional node points $\{ o^\mathcal{T}_i \}$ with spacing $1/G$ and an offset $1/(G + 2)$, meaning that $G$-level temporal resolution is composed of the node points $\{\frac{1}{G + 2}, \frac{1}{G + 2} + \frac{1}{G}, \dots, 1 - \frac{1}{G + 2}\} \}$.
As the scaling parameters $\zeta_H$ and $\zeta_G$, for spatial and temporal radial basis functions, we use $\zeta_H = \frac{1}{2.5 H}$ following \cite{nychka2015multiresolution} and $\zeta_G = \frac{|o^{\mathcal{T}}_1 - o^{\mathcal{T}}_2|}{\sqrt{2}}$ following \cite{spatiotemporal_deepkriging}. \mika{Spatial and temporal node points and radial basis functions are illustrated in Figure~\ref{fig:spatial_and_radial_aux_variables} for $H = 2$ spatial resolution, producing 4 spatial radial basis functions, and $G = 5$, producing 5 temporal radial basis functions. In practice, multiple spatial and temporal resolution levels, such as $H=(H_1, H_2) = (2, 9)$, and $G = (G_1, G_2, G_3) = (9, 17, 37)$, should be used to capture both large scale and finer dependencies.} 
%For radial basis function based iVAE, we form the auxiliary variable using spatial resolution levels \mika{$H=(H_1, H_2) = (2, 9)$}, and temporal resolution levels \mika{$G = (G_1, G_2, G_3) = (9, 17, 37)$}. 
An advantage of using radial basis functions as auxiliary variables instead of spatio-temporal segments is that by using radial basis functions, iVAE's auxiliary function provides a smooth spatio-temporal trend and variance functions, which can be used later for further analysis such as for prediction purposes. The radial basis function based iVAE is denoted as iVAEr in the rest of the paper.

\section{Simulation studies}
\label{sec:simulations}

The aim of this section is to demonstrate and compare the performances of iVAE methods using simulation studies and to discover how various types of nonstationarity in variance affect the performance. The section begins with a short review of some common procedures for generating spatio-temporal data and ways to introduce nonstationarity in it. The remainder of the section contains a large simulation study showing the unmixing performances of the iVAE methods under different types of nonstationarity scenarios, and then introduces two methods to estimate the number of latent signals. Finally, the latent dimension estimation methods are illustrated in a small simulation study.
%The rest of the section contains a small 
%simulation study of estimating the number of independent components and a larger study showing the unmixing performance of the iVAE models under different types of nonstationary data. 
All simulations can be reproduced using R 4.3.0 \cite{Rcoreteam} together with R packages fastICA \cite{fastICA}, SpaceTimeBSS \cite{SpaceTimeBSS} and NonlinearBSS. NonlinearBSS package contains R implementations of all proposed spatio-temporal iVAE variants, and is available in \url{https://github.com/mikasip/NonlinearBSS}. The simulations were executed on the CSC Puhti cluster, a high-performance computing environment.

%\klaus{The following section might be nice for a thesis intro chapter but not really relevant here. Either putting most into an appendix or right away shortening it just so that one can see how the data was generated.}

\subsection{Nonstationary spatio-temporal data generation}
\label{sec:ilsa}

Spatio-temporal data are typically composed of $n_s$ spatial locations and $n_t$ temporal points, making the total number of observations $n = n_s n_t$ usually very high. The observations are often collected regularly, for example daily or hourly, by some monitoring stations in different locations. This makes the observed data quickly very dense in time but more sparse in space. To study the properties of the models under the nature of real life spatio-temporal data, generating large datasets with various spatio-temporal covariance models is required. In the following simulations, we exploit a computationally efficient vector autoregressive process, see for example \cite{sigrist2012dynamic, papalexiou2020random, xu2018improved, yan2021vector}, and a simplified version of improved latent space approach (ILSA) \cite{xu2018improved} to generate nonstationary spatio-temporal data.

Assume spatial field at time $t=1,\dots, n_t$ to be $\bs \delta(t) = (\delta(\bs s_1, t), \dots, \delta(\bs s_{n_s}, t))^\top$, where $\bs s_1, \dots \bs s_{n_s}$ are the spatial locations in the spatio-temporal field. The vector autoregressive process can be written as
\begin{align}
\label{eq:vector_ar}
\bs \delta(t) = \sum_{r = 1}^R \rho_r \bs K_r \bs \delta(t - r) + \bs \epsilon_{\bs \delta}(t),
\end{align}
where $r = 1, \dots, R$ is an autoregressive order, $\rho_r$ is $r$th baseline autoregressive coefficient, $\bs K_r$ is a $n_s \times n_s$ spatial kernel matrix determining the change of temporal correlation with spatial locations, and $\epsilon_{\bs \delta}(t)$ is a $n_s$-dimensional noise vector with covariance $C(\epsilon_{\bs \delta}(\bs s, t), \epsilon_{\bs \delta}(\bs s', t))$ with $\bs s, \bs s' \in \{ \bs s_1, \dots, \bs s_{n_s}\}$. 

With a simplified version of ILSA, one can generate nonstationary spatio-temporal data by using vector autoregressive process (\ref{eq:vector_ar}) as formulated next. Let $\tilde{\bs s}(\bs s)=[\tilde{s}_1, \dots, \tilde{s}_d]$ be a $d$-dimensional transformation of the original coordinate $\bs s$, where the transformed coordinates $\tilde{s}_i$, $i=1,\dots,d$, are called regressors or latent coordinates. Let $d_{\bs s^i \bs s^j} = [\| s_1^1 - s_1^2\|, \|s_2^1 - s_2^2 \|]^\top$, $d_{\tilde{\bs s}^{i} \tilde{\bs s}^{j}} = [\| \tilde{s}_1^{1} - \tilde{s}_1^{2}\|, \dots, \|\tilde{s}_d^{1} - \tilde{s}_d^{2} \|]^\top$ and $V$ to be any stationary covariance function. Simplified ILSA has the formulations 
%\textcolor{red}{Try to make equations below look nicer. Matrices look somehow too large? Replace square root by $(\cdot)^{1/2}$..}
\begingroup
\renewcommand*{\arraystretch}{0.7}
\setlength{\arraycolsep}{1pt}
\begin{align}
&K_{r\{i,j\}} = 
\begin{vmatrix}
    \bs \theta_{\bs s, r} &  \\
     & \bs \theta_{\tilde{\bs s}, r} 
\end{vmatrix}^{-\frac{1}{2}}
\text{exp}\left( 
-\begin{bmatrix}
    \bs d_{\bs s_i \bs s_j} & \bs d_{\tilde{\bs s}_i \tilde{\bs s}_j}
\end{bmatrix}
\begin{bmatrix}
    \bs \theta_{\bs s, r} &  \\
     & \bs \theta_{\tilde{\bs s}, r}
\end{bmatrix}
\begin{bmatrix}
    \bs d_{\bs s_i \bs s_j} \\ 
    \bs d_{\tilde{\bs s}_i \tilde{\bs s}_j}
\end{bmatrix}
\right), \nonumber \\
&C(\epsilon_{\bs \delta}(\bs s, t), \epsilon_{\bs \delta}(\bs s', t)) = \sigma[\tilde{\bs s}(\bs s), \bs s, t] \sigma[\tilde{\bs s}(\bs s'), \bs s', t] V(Q_t) \label{eq:ILSA} \\
&\text{where } Q_t = \left(
\begin{bmatrix}
     \bs d_{\bs s_i \bs s_j} & \bs d_{\tilde{\bs s}_i \tilde{\bs s}_j}
\end{bmatrix}
\begin{bmatrix}
    \bs \theta_{\bs s} &  \\
     & \bs \theta_{\tilde{\bs s}}
\end{bmatrix}
\begin{bmatrix}
    \bs d_{\bs s_i \bs s_j} \\ 
    \bs d_{\tilde{\bs s}_i \tilde{\bs s}_j}
\end{bmatrix}\right)^\frac{1}{2} \nonumber
\end{align}
\endgroup
and $\bs \theta_{\bs s, r}=\text{diag}(\theta_{s_1,r}, \theta_{s_2, r}), \bs \theta_{\tilde{\bs s}, r}=\text{diag}(\theta_{\tilde{s}_1,r}, \dots, \theta_{\tilde{s}_d, r}), \bs \theta_{\bs s}=\text{diag}(\theta_{s_1}, \theta_{s_2})$ and $\bs \theta_{\tilde{\bs s}}=\text{diag}(\theta_{\tilde{s}_1,r}, \dots, \theta_{\tilde{s}_d, r})$ are diagonal matrices giving scaling parameters for the spatial coordinates and for the latent coordinates. The function $Q_t$ transforms the original coordinates based on the scaling parameters and the latent coordinates. By using this approach, one can easily introduce complex, nonstationary and nonseparable covariance structures through latent coordinates $\tilde{\bs s}$, time varying spatial kernel matrices $\bs K_r$ and nonstationary variance function $\sigma$.
In the following simulations, we are mainly interested in having nonstationarity in the variance as that is required for the identifiability of the latent components.

\subsection{Finite sample efficiencies}

%In this section four different iVAE configurations, \mika{regular VAE}, symmetric FastICA (FICA) \cite{hyvarinen1999fast} with hyperbolic tangent nonlinearity and STBSS are compared using simulated spatio-temporal data. \mika{Although FICA is not developed for spatio-temporal data or nonlinear mixing, it is considered as a linear baseline for data with nonstationary variance. STBSS is developed for stationary spatio-temporal data and linear mixing, and is considered as a spatio-temporal baseline. 
%Even though there are multiple deep learning based attempts for nonlinear BSS in the literature \cite{ANSARI2023126895}, most of them are not identifiable and focus on acoustic data, where only serial dependence is present making them nonoptimal for spatio-temporal data. Nevertheless, we consider regular VAE as an unidentifiable deep learning baseline.
%} 

\mika{In this section, four different iVAE configurations--regular VAE, symmetric FastICA (FICA) \cite{hyvarinen1999fast} with hyperbolic tangent nonlinearity, and STBSS--are compared using simulated spatio-temporal data. Although FICA is not designed for spatio-temporal data or nonlinear mixing, it is included as a linear baseline for data with nonstationary variances. STBSS, developed for stationary spatio-temporal data and linear mixing, serves as a spatio-temporal baseline.

While there are several deep learning-based approaches for nonlinear BSS in the literature, see \cite{ANSARI2023126895} for a recent review, most lack identifiability and focus on acoustic data, which primarily exhibits serial dependence, making them suboptimal for spatio-temporal data. Nonetheless, we include regular VAE as an unidentifiable deep learning baseline.}

\mika{The aim of the simulations is to identify how the proposed iVAE methods perform as compared to other existing methods under various types of nonstationary spatio-temporal data, and how the type of nonstationary affects the performance.} To identify how the reduction of either temporal or spatial observations affect the performance of the algorithms, we consider three sample sizes composed of $n_s$ spatial locations and $n_t$ temporal observations for each spatial location. The sample dimensions considered are $(n_s, n_t) = (150, 300)$, $(n_s, n_t) = (50, 300)$ and $(n_s, n_t) = (150, 75)$ yielding $n=45000$, $n=15000$ and $n=11250$ observations, respectively. We generate the latent data $\bs z$ according to six different simulation settings. In some settings the nonstationarity is introduced only in time, in some settings only in space, and in some settings both in space and in time. In each simulation setting, $n_s$ spatial locations $\bs s_i$ are sampled uniformly in spatial domain $\mathcal{S} = [0,1] \times [0,1]$, and for each spatial location $\bs s_i$, $n_t$ observations $\bs x(\bs s_i, t)$ are generated. The true latent dimension is $P=5$ and the dimension of the observations is $S=8$. Every setting is repeated $500$ times for each sample size and for each algorithm. Finally, each trial is repeated using three increasingly nonlinear mixing functions as described hereafter. The first three simulation settings are more simple ones, followed by three more complex ones which utilize the ILSA framework. The simulation settings and the mixing procedure are defined in the following. \mika{After introducing the data generation of the settings, the motivation behind each setting is carefully explained.}

\textbf{Setting 1.} The latent spatio-temporal field consists of three clusters in space and five segments in time yielding 15 spatio-temporal clusters, each of which has their own unique diagonal covariance matrix and unique mean vector. For $k$th cluster, $k=1,\dots, 15$, the covariance matrix is given as $\bs C_k = \text{diag}(\sigma_{1,k}, \dots, \sigma_{5,k})$, where $\sigma_{i,k}, \sim \text{Unif}(0.1,5)$ and unique mean vector is given as $\bs \mu_k = (\mu_{1,k}, \dots, \mu_{5,k})^\top$, where $\mu_{i,k} \sim \text{Unif}(-5,5)$, $i=1,\dots,5$.

\textbf{Setting 2.} The latent spatio-temporal field consists of $10$ segments in time. The latent components are simulated by generating first Gaussian spatial data with Matern covariance function using unique parameters $(\nu_i, \phi_i)$ for each component $i=1,\dots,5$, and then adding Gaussian iid data with unique covariance matrix and mean vector for each time segment. The Matern parameters are $(\nu_1, \phi_1) = (0.5, 0.30)$, $(\nu_2, \phi_2) = (0.1,0.25)$, $(\nu_3, \phi_3)=(1, 0.35)$, $(\nu_4, \phi_4)=(2, 0.20)$, $(\nu_5, \phi_5)=(0.25, 0.15)$ and the parameters for the time segment $k=1, \dots, 10$ are $\bs \mu_k = (\mu_{1,k}, \dots, \mu_{5,k})^\top$, where $\mu_{i,k} \sim \text{Unif}(-0.3,0.3)$, and $\bs \Sigma_k = \text{diag}(\sigma_{1,k}, \dots, \sigma_{5,k})$, where $\sigma_{i,k}, \sim \text{Unif}(0,0.4)$, $i=1,\dots,5$.

\textbf{Setting 3.} The latent spatio-temporal field consists of five clusters in space and follows AR1 model. In $k$th cluster, $k=1, \dots, 5$, the latent components $z_i$, $i = 1, \dots, 5$, are generated as 
\begin{align*}
    z_i(\bs s, t + 1) &= \rho_{i,k} z_i(\bs s, t) + \epsilon_{i,k,t}, \\
    \epsilon_{i,k,t} &\sim N(\mu_{i,k}, \sigma_{i,k}),
\end{align*}
where $t=1,\dots, n_t - 1$ and $z_i(\bs s, 1) \sim N(\mu_{i,k}, \sigma_{i, k})$. Each cluster has unique parameters $\bs \rho_k = (\rho_{1,k}, \dots \rho_{5,k})^\top$, $\bs \mu_k = (\mu_{1,k}, \dots \mu_{5,k})^\top$ and $\bs \sigma_k = (\sigma_{1,k}, \dots \sigma_{5,k})^\top$ generated as $\rho_{i,k} \sim \text{Unif}(0.05, 0.95)$, $\mu_{i,k} \sim \text{Unif}(-1, 1)$ and $\sigma_{i,k} \sim \text{Unif}(0.1, 5)$.

%\textbf{Setting 3} The latent spatio temporal field contains zero-mean Gaussian data with varying diagonal covariance matrices over 5 segments in time with the following univariate spatio temporal covariance model:
%\begin{align}
%C(h, \tau; \gamma, \nu, \phi, a_1, a_2, a_3) &= a_1 C_{\text{AR1}}(\tau; \gamma) C_{\mathcal{M}}(h; \nu, \phi) + a_2 C_{\text{AR1}; \gamma}(\tau) + a_3 C_{\mathcal{M}}(h; \nu, \phi), \\
%C_{\text{AR1}}(\tau; \gamma) &= \gamma^\tau, \\
%C_{\mathcal{M}}(h; \nu, \phi) &= \frac{1}{2^{\nu-1}\Gamma(\nu)}\left(\frac{h}{\phi}\right)^\nu K_\nu\left(\frac{h}{\phi}\right),    
%\end{align}
%where $K$ is a modified Bessel function of second kind, $h = \| \bs s_i - \bs s_j \|$ and $\tau = |t_i - t_j|$. The temporal domain is divided randomly into 5 segments, and in each segment the data is generated with unique covariance matrix $\bs \Sigma_i = \text{diag}(\sigma_{1,i}, \dots, \sigma_{5,i})$, where $\sigma_{j,i}, \sim \text{Unif}(0.1,5)$. The parameters $(\gamma_i, \nu_i, \phi_i, a_{1,i}, a_{2,i}, a_{3,i})$ for ICs 1-5 are provided in Table \ref{tab:setting3_params}.

\textbf{Settings 4-6.} The latent spatio-temporal field is generated using ILSA framework. Each setting has the same highly nonstationary covariance structure. In addition, Setting 4 has a variance $\sigma$ changing in space, Setting 5 has a variance changing in time, and Setting 6 has a variance changing both in space and in time. The latent coordinates $\tilde{\bs s} = (\tilde{s}_1, \tilde{s}_2)^\top$ are transformed from the spatial coordinates $\bs s=(s_1, s_2)^\top$ by using a swirl-like coordinate transformation according to \cite{papalexiou2021advancing} given by
\begin{align*}
    \tilde{s_1} &= (s_1 - s_1^*) \text{ cos} \Biggl( \eta \text{ exp} \left( -(\frac{h^*}{b_{\text{swirl}}})^2 \right) \Biggl) - (s_2 - s_2^*) \text{ sin} \Biggl( \eta \text{ exp} \left( -(\frac{h^*}{b_{\text{swirl}}})^2 \right) \Biggl) + s_1^*,  \\
    \tilde{s_2} &= (s_1 - s_1^*) \text{ sin} \Biggl( \eta \text{ exp} \left( -(\frac{h^*}{b_{\text{swirl}}})^2 \right) \Biggl) - (s_2 - s_2^*) \text{ cos} \Biggl( \eta \text{ exp} \left( -(\frac{h^*}{b_{\text{swirl}}})^2 \right) \Biggl) + s_2^*, 
\end{align*}
where $\bs s^*=(s_1^*, s_2^*)$ is the center point of the deformation, $h^*=\|\bs s - \bs s^*\|$ is Euclidean distance between the original location and the center point, $\eta$ is a rotation angle, and $b_{\text{swirl}}$ is a scaling parameter controlling the magnitude of the swirl. Each latent component has their own set of deformation parameters. The stationary covariance function $V$ in (\ref{eq:ILSA}) is the Matern covariance function 
%\textcolor{red}{In 2 you talk about Matern correlation structure, please unify} 
with parameters $(\nu_i, \phi_i)$ for all Settings 4-6. The deformation parameters, ILSA parameters and Matern parameters mutual for Settings 4-6 are given in Table~\ref{tab:ilsa_params}. The autoregressive order is $R = 1$ for all settings.

In Setting 4, we have $\sigma[\tilde{\bs s}(\bs s), \bs s, t] = \text{exp}(\theta_{\sigma_s}^i(\tilde{s}_1 - 0.5))$, where $\theta_{\sigma_s}^i$ is the scaling parameter of variance in space for $i$th latent component. This means that the variance of the latent fields vary in space based on the first latent coordinate. The variance scaling parameters for the latent components $z_i$, $i=1,\dots, 5$, are $\theta_{\sigma_s}^1=1$, $\theta_{\sigma_s}^2=2$, $\theta_{\sigma_s}^3 = 3$, $\theta_{\sigma_s}^4 = -1$ and $\theta_{\sigma_s}^5 = -2$.

In Setting 5, the variances of the latent fields are changing in time. We set $\sigma[\tilde{\bs s}(\bs s), \bs s, t] = \text{exp}(\text{sin}((t + \theta_{\sigma_{t1}}^i) + \theta_{\sigma_{t2}}^i) / 2)$, where $\theta_{\sigma_{t1}}^i$ and $\theta_{\sigma_{t2}}^i$ are variance coefficient and variance scaling parameter in time for $i$th latent component. The parameters  $(\theta_{\sigma_{t1}}^i, \theta_{\sigma_{t2}}^i)$ for the latent components $z_i$, $i=1,\dots,5$, are $(\theta_{\sigma_{t1}}^1, \theta_{\sigma_{t2}}^1) = (50, 0.1)$, $(\theta_{\sigma_{t1}}^2, \theta_{\sigma_{t2}}^2) = (0, 0.05)$, $(\theta_{\sigma_{t1}}^3, \theta_{\sigma_{t2}}^3) = (100, 0.005)$,
$(\theta_{\sigma_{t1}}^4, \theta_{\sigma_{t2}}^4) = (20, 0.01)$ and
$(\theta_{\sigma_{t1}}^5, \theta_{\sigma_{t2}}^5) = (10, 0.03)$.

In Setting 6, the variances of the latent fields are changing in space and in time. We set $\sigma[\tilde{\bs s}(\bs s), \bs s, t] = \text{exp}(\theta_{\sigma_s}^i(\tilde{s}_1 - 0.5) + \text{sin}((t + \theta_{\sigma_{t1}}^i) + \theta_{\sigma_{t2}}^i) / 2)$. The parameters $\theta_{\sigma_{s}}^i, \theta_{\sigma_{t1}}^i, \theta_{\sigma_{t2}}^i$ for the latent fields $z_i$, $i = 1,\dots, 5$, are identical as in Settings 4 and Setting 5. 

Setting 1 has the simplest latent fields by having a diagonal spatio-temporal covariance for each latent field. The variance and mean are changing explicitly between the spatio-temporal clusters as is assumed for segmentation based iVAE. This setting is a spatio-temporal variant of the simulation setting used in time series context in \cite{Khemakhem2020, HyvarinenMorioka2016}, where the latent components had multiple temporal segments with unique mean and/or variance parameters. Settings 2 and 3 are still relatively simple with no spatio-temporal interaction in the latent fields. Setting 2 is used to compare performances in cases where latent fields are stationary in space, but variance is changing over time. Setting 3 illustrates a scenario where the latent fields are stationary in time, but the variance is changing over the clusters in space. By having less variability in the variance, \mika{Settings 2 and 3} should be less optimal for iVAE. Settings 4-6 have latent fields with a complex spatio-temporal covariance model and strong spatio-temporal interaction. The variance is not changing explicitly over segments, but instead through a nonstationary covariance function. In Setting 4, the latent fields have smoothly changing nonstationary variance in space, but the variance is stationary in time, and in Setting 5, the variance is nonstationary in time, but stationary in space. Setting 6 introduces nonstationarity in variance both in space and in time. With Settings 4-6 the aim is thus to find out how the presence of nonstationarity in variance affects the performances of iVAE methods in settings with more realistic and more complex spatio-temporal structures.

\begin{table*}[]
    \centering
    \caption{The ILSA parameters and coordinate deformation parameters for Settings 4-6.}
    \resizebox{1\textwidth}{!}{
    \begin{tabular}{lllllllllll}
    \hline
         & $\bs \theta_{\bs s, l}$ & $\bs \theta_{\tilde{\bs s}, l}$ & $ \bs \theta_{\bs s}$ & $\bs \theta_{\tilde{\bs s}}$ & $\rho_1$ & $\bs s^*$ & $b_{\text{swirl}}$ & $\eta$ & $\nu$ & $\phi$ \\
             \hline
         IC1 & $(6, 4)$ & $(7, 7)$ & $(0.2, 0.7)$ & $(0.7, 0.2)$ & 0.9 & $(0.5, 0.5)$ & $0.7$ & $1.8\pi$ & 0.25 & 0.5 \\
         IC2 & $(3, 6)$ & $(4, 7)$ & $(0.7, 0.2)$ & $(0.25, 0.5)$ & 0.8 & $(0.7, 0.7)$ & $0.4$ & $1.2\pi$ & 0.2 & 0.9 \\
         IC3 & $(3, 3)$ & $(6, 3)$ & $(0.5, 0.5)$ & $(0.7, 0)$ & 0.7 & $(0.3, 0.3)$ & $0.2$ & $2\pi$ & 0.05 & 1.5 \\
         IC4 & $(7, 3)$ & $(2, 6)$ & $(0.2, 0.4)$ & $(0.3, 0.7)$ & 0.6 & $(0.7, 0.3)$ & $1$ & $0.5\pi$ & 0.1 & 0.25 \\
         IC5 & $(2, 1)$ & $(6, 2)$ & $(0.3, 0.3)$ & $(0, 0.7)$ & 0.5 & $(0.3, 0.7)$ & $0.9$ & $0.9\pi$ & 0.15 & 1 \\
         \hline
    \end{tabular}}
    \label{tab:ilsa_params}
\end{table*}

\textbf{Nonlinear mixing functions.} The mixing function $f_L$ is generated using multilayer perceptron (MLP) following \cite{Khemakhem2020, HyvarinenMorioka2016, Hyvarinen2019}. Here $L$ denotes the number of mixing layers used in MLP. To obtain an injective and differentiable mixing function, each layer of MLP has $S=8$ hidden units with the activation function $\omega_i$ being either linear or exponential linear unit (ELU). The matrices $\bs B_i$, $i=1,\dots,L$, in the mixing procedure are normalized to have unit length row and column vectors to guarantee that none of the independent components vanish in the mixing process. The mixing function $f_L$ is defined as
\begin{align*}
    \bs f_L(\bs z) = \begin{cases}
        \omega_L(\bs B_L \bs z),\quad L = 1, \\
        \omega_L(\bs B_L \bs f_{L-1}(\bs z)),\quad L \in \{2,3,\dots\},
    \end{cases}
\end{align*}
where $\bs B_1$ is a $8 \times 5$ matrix and the other matrices, $\bs B_i$, $i \neq 1$, are $8 \times 8$ matrices.
In simulations we use linear activation $\omega_L(x)=x$ for the last layer and ELU activation 
\begin{align*}
\omega_i(x)=\begin{cases}  
    x,\quad x \geq 0, \\
    \text{exp}(x) - 1,\quad x < 0,
\end{cases}
\end{align*} $i = 1,\dots, L-1$, for the other layers. By this procedure, with the number of mixing layers $L=1$, we obtain $S=8$ linear mixtures of the independent components. When the number of mixing layers increase, the mixtures become increasingly nonlinear. In simulations, we consider three different mixing functions with the number of mixing layers $L= 1, 3, 5$.

\begin{figure*}
  \centering
  \includegraphics[width=1\textwidth]{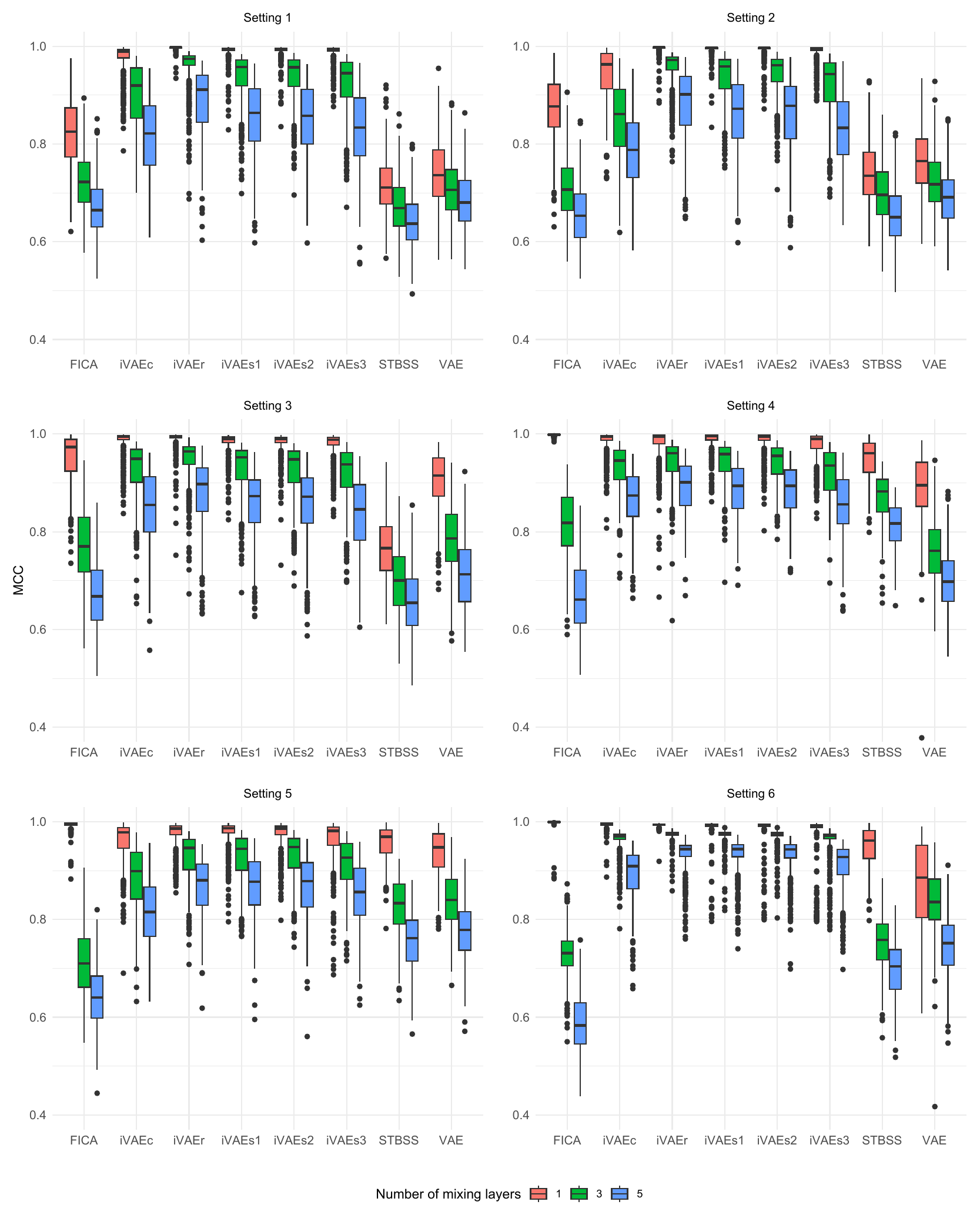}
  \caption{Mean correlation coefficients of 500 trials for Settings~1-6 for sample size with the number of spatial locations $n_s = 150$ and the number of temporal observations $n_t = 300$.}
  \label{fig:sim_all_s150t300}
\end{figure*}

\textbf{Model specifications.} All iVAE models are set up with encoder, decoder and auxiliary function with three hidden layers in each. The hidden layers consist of 128 neurons and leaky rectified linear unit (leaky ReLU) activation \cite{leaky_relu}. iVAEs1, iVAEs2 and iVAEs3 use $4 \times 4$ spatial segmentation, resulting a grid of $m_S = 16$ equally sized squares. The temporal segmentation is done by dividing the temporal domain to equally sized segments of length 5. This results the number of temporal segments $m_T = 60$ when $n_t = 300$ and $m_T = 15$ when $n_t = 75$. For iVAEr, we use spatial resolution levels $H = (2, 9)$, and temporal resolution levels $G = (9, 17, 37)$. The iVAE models are trained for 60 epochs when $(n_s, n_t) = (150, 300)$, for 120 epochs when $(n_s, n_t) = (50, 300)$ and for 150 epochs when $(n_s, n_t) = (150, 75)$. The number of epochs is increased when the sample size is decreased, as the number of training steps in each epoch is lower for the smaller sample size. For all sample sizes, the number of epochs are selected large enough to guarantee that the training converges. All iVAE models use learning rate of 0.001 with polynomial decay of second-order over 10000 training steps, where the learning rate after the first 10000 training steps is 0.0001. \mika{VAE uses similar parameters as iVAE, but it does not use any auxiliary data or have an auxiliary function.}
%For FICA, hyperbolic tangent nonlinearity is used with convergence tolerance $10^{-4}$. 
The STBSS model is fitted with multiple different kernel settings, and the best one is selected, which is having two spatial ring kernels $(0, 0.15)$ and $(0.15, 0.3)$ and time lag of 1. For more about STBSS and its kernel settings, see \cite{MuehlmannDeIacoNordhausen2022}.

\textbf{Performance index.} To measure the performance of the methods, the mean correlation coefficient (MCC) is used following the previous studies, e.g., \cite{HyvarinenMorioka2017, Hyvarinen2019, HalvaHyvarinen2020, SipilaNonlinearSBSS}. MCC is a function of the correlation matrix $\bs \Omega = Cor(\bs z, \hat{\bs z})$ between the true latent components $\bs z$ and the estimated ones $\hat{\bs z}$. MCC is calculated as
\begin{align}
    \text{MCC}(\bs \Omega) = \frac{1}{P}\sup_{\bs P \in \mathcal{P}} \text{tr}(\bs P\, \text{abs}(\bs \Omega)),
\end{align}
where $\mathcal{P}$ is a set of all possible $P \times P$ permutation matrices, $\text{tr}(\cdot)$ is the trace of a matrix and $\text{abs}(\cdot)$ denotes taking elementwise absolute values of a matrix. MCC gets values in range $[0, 1]$, where the optimal value 1 is obtained when the estimated sources are correlated perfectly up to their signs with the true sources.

\textbf{Results.} The simulation results are provided in Figure~\ref{fig:sim_all_s150t300} for $(n_s, n_t) = (150, 300)$ and in Figures~\ref{fig:sim_all_s50t300} and \ref{fig:sim_all_s150t75} in the \ref{sec:appendix} for $(n_s, n_t) = (50, 300)$ and $(n_s, n_t) = (150, 75)$, respectively. Based on the results, it is clear that only the iVAE methods are capable of recovering sources through nonlinear unmixing environment. The performances of iVAE methods are better, when the sample size grows, although the differences are small in some settings. The performance of iVAEc is slightly worse than the performances of the other iVAE methods in every setting. FICA performs well in the linear mixing settings, when the latent fields do not contain trend in mean. In nonlinear settings, its performance drops dramatically in all simulation settings. Similar behaviour is present for STBSS, although the performance in zero mean settings does not reach FICA. This is not surprising as STBSS is developed for stationary spatio-temporal random fields. \mika{VAE performs poorly in almost all settings, which is expected as the model is not identifiable.}

In Settings 1-3 all iVAE methods outperform FICA, STBSS and VAE. In these settings, the best performing method is iVAEr, followed by iVAEs1, iVAEs2 and iVAEs3, respectively. They all perform very well under the linear mixing, but when the number of mixing layers is increased, iVAEr outperforms the three other methods. iVAEs1 and iVAEs2 have very similar performance, and they perform slightly better than iVAEs3. The performance of iVAEc is worse than performances of other iVAE methods, especially in Setting 2. FICA performs relatively well in Setting 3 under the linear mixing, but the performance is poor in other settings. STBSS fails to recover the latent fields in Settings 1-3 even under the linear setting. \mika{VAE fails in Settings 1 and 2, but performs moderately in Setting 3 under linear mixing.}

In Settings 4-6, the best method under linear mixing is FICA, but its performance drops considerably in nonlinear settings. In case of nonlinear mixing, the best methods are iVAEs1 and iVAEs2 and iVAEr followed by iVAEs3 and iVAEc, in the order from best to worst. iVAEs1, iVAEs2 and iVAEr perform nearly as well as FICA in linear setting and keep up their good performance also in nonlinear settings. STBSS performs rather well under linear mixing, but is still worse than FICA and iVAE variants. \mika{VAE has slightly lower performance than STBSS under linear mixing, but it also fails under nonlinear mixing.} In Setting 5, when $n_t = 75$, the performances of all the methods drop considerably. This is probably due to the fact that in Setting 5, the variance is varying less over the whole temporal domain when the number of time points is lower. The best performing methods, when $n_t=75$, are iVAEs1 and iVAEs2 in both linear and nonlinear cases. In Setting 6, the performance of iVAE methods drop only slightly when the number of mixing layers is increased, especially when the sample size is high. The best performing methods for nonlinear mixing environment are iVAEr, iVAEs1 and iVAEs2.

In general, if the variability of the variance remains the same, increasing the sample size improves the results only a little. The differences in MCC between the smallest and the largest sample sizes are between 0.005 and 0.03 for all iVAE methods and all settings except the Setting 5, which has lower MCCs when $n_t=75$. Increasing the variability of the variance has stronger impact, which is evident when the results of Settings 4, 5 and 6 are compared. The average MCCs are higher and the results are more consistent for Setting 6 than for Settings 4 and 5. Based on the results, the best three models are iVAEr, iVAEs1 and iVAEs2. Compared to segmentation based iVAE, iVAEr has an advantage of estimating a smooth spatio-temporal trend and variance functions, which are provided by the auxiliary function. These can be useful for further analysis or for prediction purposes, as is demonstrated in Section~\ref{sec:case_study}. Taking the previous facts into account, we consider the best method to be iVAEr.

%\mika{\textbf{Radial basis function parameter sensitivity.} To study the sensitivity of iVAEr to the choice of resolution levels to form the radial basis functions, we perform a small additional simulation by replicating of Setting 6 with $n_s=150$ and $n_t = 300$ using iVAEr with three different radial basis function parameter settings. The first setting has resolution levels $H=(2)$ and $G=(9)$ producing the total of 13 basis functions. The second setting is the one used in the previous simulations, $H=(2,9)$ and $G=(9,17,39)$ producing 150 basis functions. The third setting has $H=(2,9,17)$ and $G=(9,17,39,99)$ producing 538 basis functions. The results are provided in Figure~\ref{fig:radial_param_sensitivity} in Appendix~\ref{sec:appendix}. Based on the results, too few radial functions should not be used as the first setting, $H=(2), G=(9)$, has lower performance compared to the settings with more radial basis functions. The other two settings provided very nearly identical results, which hints that the algorithm is not sensitive against the choice of resolution levels as long as the number of spatial and temporal basis functions are not too small. As a guideline we suggest starting with the resolution levels $H=(2, 9)$ and $G=(9, 17, 39)$ for obtaining good performance with fairly low computational complexity. Notice also, that it is beneficial to use multiple varying spatial and temporal resolution levels (e.g. $H=(2,9)$ instead of $H=(10)$) to ensure that both large scale dependencies and finer dependency structures are captured.}

\mika{\textbf{Radial basis function parameter sensitivity.} To examine the sensitivity of iVAEr to the choice of resolution levels for forming radial basis functions, we conduct an additional small-scale simulation by replicating Setting 6 with $n_s=150$ and $n_t=300$, using iVAEr with three different radial basis function parameter configurations. The first configuration uses resolution levels $H=(2)$ and $G=(9)$, resulting in a total of 13 basis functions. The second configuration, used in previous simulations, has $H=(2,9)$ and $G=(9,17,39)$, producing 150 basis functions. The third configuration uses $H=(2,9,17)$ and $G=(9,17,39,99)$, producing 538 basis functions.

The results, presented in Figure~\ref{fig:radial_param_sensitivity} in \ref{sec:appendix}, indicate that using too few radial basis functions leads to lower performance, as seen with the first setting ($H=(2), G=(9)$). The other two settings produced nearly identical results, suggesting that the algorithm is not highly sensitive to the choice of resolution levels, provided the number of spatial and temporal basis functions is sufficient. As a general guideline, we recommend starting with resolution levels $H=(2,9)$ and $G=(9,17,39)$ for good performance with relatively low computational complexity. Additionally, it is advantageous to use multiple varying spatial and temporal resolution levels (e.g., $H=(2,9)$ instead of $H=(10)$) to capture both large-scale dependencies and finer dependency structures.}

%\begin{figure}
%  \centering
%  \includegraphics[width=1\textwidth]{nonlinear_mixing_and_preds.pdf}
%  \caption{Illustration how the nonlinearity increases when more mixing layers are applied (number of layers increases from top to bottom) to data with $P=5$ independent components and number of observations $S=5$. Red lines are estimated mixing function by iVAE, when the data is generated according to setting 1. The x-axis has the values of the first independent component and the y-axis has the values of the first observed mixture, when the other independent components have $z_i=0$, $i=2,\dots, P$. }
%  \label{fig:mixing_f_pred}
%\end{figure}

\subsection{Latent dimension estimation}

\mika{
In the previous section, we demonstrated through simulations that when the number of latent components $P$ is known, iVAE effectively recovers these components. However, in practical applications, the true number of latent components is often unknown and must be estimated. Accurately determining the dimension is critical; too few components may lead to the omission of vital information, flawed interpretations, and erroneous predictions. Conversely, too many components can complicate interpretation and typically result in overly noisy predictions. One of the reasons for the success of BSS is that it generally simplifies interpretability and enhances predictions, which is particularly valuable in handling complex data such as in spatio-temporal contexts.

Note that the estimation of the number of components in BSS models with linear mixing has only been considered recently. These approaches typically handle iid data, or temporal or spatial data, but, to the best of our knowledge, they have not yet been applied to spatio-temporal data. Moreover, these methods often rely on eigenvalues of specially constructed scatter matrices, which precludes their extension to nonlinear scenarios. For more details on these approaches, see \cite{LuoLi2016, LuoLi2020, VirtaNordhausen2021, NordhausenTaskinenVirta2022, YiNordhausen2023, MuehlmannBachocNordhausenYi2024, RadojicicNordhausen2024} and the references therein.

To facilitate the estimation of the number of components within our framework, this section proposes two alternatives: a visual procedure to select the latent dimension and a more formal latent dimension estimation method. We illustrate these methods using simulations. In all simulations, we adopt two setups: one where $P=5$ and $S=8$, as in the previous simulations, and another where $P=10$ and $S=15$. In both setups, the latent components are generated as in Setting~6 of the previous section with randomly generated parameters from appropriate uniform distributions, and the observed data is generated using the same mixing process as in the previous simulations. For these simulations, we employ only iVAEr with the same parameters as in the previous section and train it for 60 epochs in each trial.}

\begin{figure}
  \centering
  \subfloat[]{
  \includegraphics[width=0.45\textwidth]{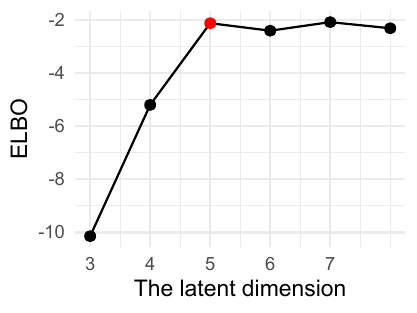}}
  \subfloat[]{
  \includegraphics[width=0.45\textwidth]{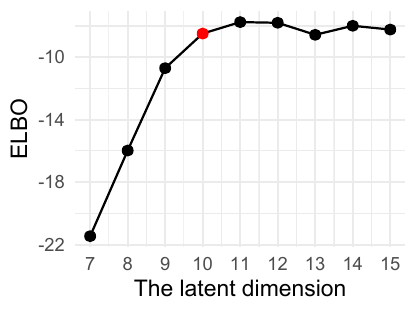}}
  \caption{The lower bounds of the data log likelihoods plotted for different latent dimensions. The true latent dimension is marked with red. The first figure (a) has the setting with $P=5$ and $S=8$, and the second one (b) has the setting with $P=10$ and $S=15$. In second figure, the latent dimensions $R=2,\dots,6$ are cut out to make the differences more visible for larger latent dimensions.}
  \label{fig:sim_ic_est}
\end{figure}

\textbf{Visual knee point detection.} As iVAE is capable of estimating injective mixing functions, it is possible to fit multiple iVAE models using the latent dimension $R=2, \dots, S$ and compare ELBOs of fitted models. The likelihood has its maximum when the true latent components are found, and hence, ELBO tends to increase rapidly when $R<P$ and stay approximately the same when $R \geq P$. Because of this behaviour, when ELBOs are plotted against the selected latent dimensions $R=2,\dots, S$, a ``knee'' point is visible at the point of the correct latent dimension. The knee point detection method based on ELBO was also discussed in \cite{Khemakhem2020}. The knee point behaviour is illustrated in Figure~\ref{fig:sim_ic_est} for the both setups. Here, the number of mixing layers $L=3$ were used.

\textbf{AIC based on ELBO.} A natural approach for automatic model selection is to compare Akaike information \mika{criteria} (AIC) between different models. In case of latent dimension selection, we are only interested in the latent dimension $R$, and treat the other parameters of the model as nuisance. Hence, we compare the profile AICs (pAIC, see e.g. \cite{profileAIC}) calculated as
%\textcolor{red}{I don't quite understand why $R$ is in the formula as "number of parameters"?}
\begin{align*}
    \text{pAIC} = -2 \text{log}(\mathcal{L}(R|\bs x, \bs u; \hat{\bs \theta}_R)) + 2R,
\end{align*}
where $\mathcal{L}(R|\bs x, \bs u; \hat{\bs \theta}_R)$ is the profile likelihood in which all the parameters $\bs \theta$ excluding the latent dimension $R$ have been profiled out. The data log likelihood is intractable for the generative model (\ref{eq:iVAE_generative_model}) which is why the we cannot use pAIC directly. However, if the profile log likelihood $\text{log}(\mathcal{L}(R|\bs x, \bs u; \hat{\bs \theta}_R))$ is replaced with profiled ELBO, we obtain an upper bound for pAIC:
\begin{align*}
    \text{uAIC} = -2 \text{ELBO}(R|\bs x, \bs u; \hat{\bs \theta}_R)) + 2R, 
\end{align*}
which can be used for model selection. Similarly as for regular AIC, the latent dimension which produces the lowest uAIC value is selected. To demonstrate the method, we simulated 100 datasets from both setups, $(P, S) = (5, 8)$ and $(P, S) = (10, 15)$, and then fitted iVAEr for each possible latent dimension $R=2, \dots, S$ and selected the latent dimension based on uAIC. The proportions of the selected dimensions are provided in Table~\ref{tab:sim_ic_est} for both setups.

\begin{table}[ht]
\centering
\caption{The proportions of the estimated latent dimensions $R$ for setups where $P=5$, $S=8$ (a) and $P=10$, $S=15$ (b). Each setup was repeated 100 times for each number of mixing layers $L=1,3,5$. The latent dimensions were estimated using the AIC based method.}
\subfloat[]{\begin{tabular}{rrr}
  \hline
 L & $R=4$ & $R=5$ \\
  \hline
1 & 0.00 & 1.00 \\
  3 & 0.00 & 1.00 \\
  5 & 0.03 & 0.97 \\
   \hline
\end{tabular}}
\hspace{10px}
\subfloat[]{
\begin{tabular}{rrrr}
  \hline
 L & $R=9$ & $R=10$ & $R=11$ \\
  \hline
1 & 0.00 & 0.99 & 0.01 \\
  3 & 0.00 & 0.99 & 0.01 \\
  5 & 0.07 & 0.92 & 0.01 \\
  \hline
\end{tabular}}
\label{tab:sim_ic_est}
\end{table}

In both setups, AIC based method estimated well the latent dimension for each mixing function. When the number of mixing layers is $L=1$ or $L=3$, the true latent dimension was obtained nearly every time (100 out of 100 times for setup with $(P, S)=(5, 8)$ and 99 out of 100 times for setup with $(P, S)=(10, 15)$). When $L=5$, the method underestimated the latent dimension in three trials out of 100. In setup with $(P, S)=(10, 15)$, the method overestimated the latent dimension once for each mixing function. Based on the results, uAIC can thus be seen as a promising metric for automatic latent dimension selection.

\section{Real data example}
\label{sec:case_study}

\begin{figure}
  \centering
  \includegraphics[width=0.8\textwidth]{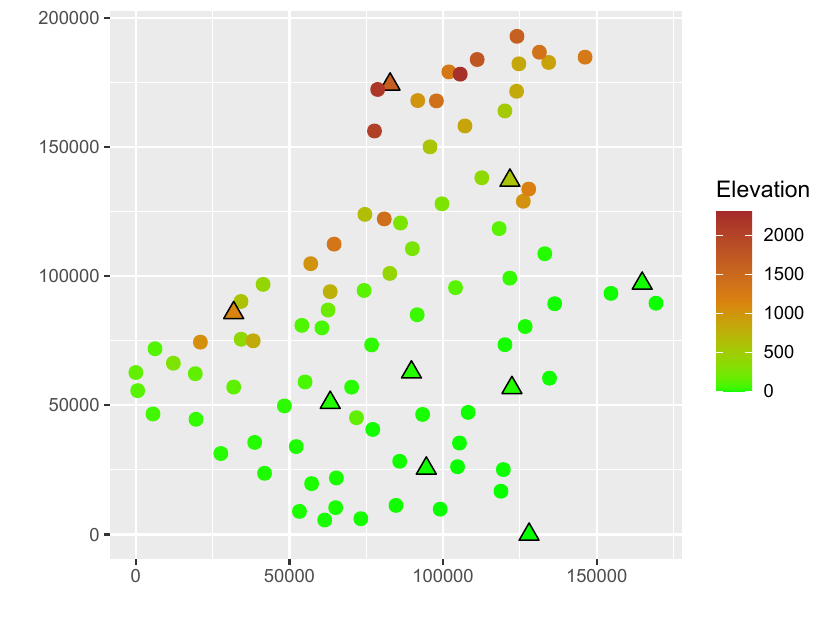}
  \caption{The spatial locations of meteorological stations in Veneto region. The validation stations are marked as triangles. The colors of the point represent the elevation differences of the stations.
%\textcolor{red}{Should we instead of 3D figure give 2D, where colors indicate the elevation? Both plots could then be bigger. Or actually only one figure (with circles and triangles) would be needed.}
}
  \label{fig:stations}
\end{figure}

Due to the best overall performance of the proposed radial basis function based spatio-temporal iVAE method, iVAEr, we now demonstrate how to apply it using a meteorological dataset collected from Veneto region in Italy. The data are collected over 23 years (2000-2022) from 101 different meteorological stations. The data consist of weekly data of evapotranspiration level (ET\_0, in mm), minimum and maximum temperature (T\_min and T\_max, respectively, in $^\circ\text{C}$), minimum and maximum humidity (H\_min and H\_max, respectively, in \%), the average wind velocity (m/s) and precipitation (mm). Therefore, we have $S = 7$. We remove 11 stations and all time points before week 28 of year 2005 to have a as little missing values as possible. After this, we have data on $90$ stations and $892$ time points. The remaining 133 rows with missing values are imputed using CUTOFF method \cite{feng2014cutoff}, which is designed for spatio-temporal imputation. We perform a log transformation $\text{log}(x + 1)$ to precipitation level to make its distribution less skew. The goal is then to find a latent representation for the data, interpret the latent components and predict the observed variables to future and to new locations by using iVAE preprocessing. We select nine random stations and all time points from beginning of May 2022 for validation purposes. As a result, we have $n_s = 81$ and $n_t = 874$ in the training data. The validation stations are presented as triangles in Figure~\ref{fig:stations} together with the rest of the stations.

In the Veneto region, the elevation of the meteorological stations vary a lot; the maximum altitude is in the north (i.e. Dolomites mountain area) and the minimum altitude can be recognised in  the south-east (i.e. Venetian plan). Hence, in this application we consider the spatial dimension $D=3$, which is composed of the X coordinate, the Y coordinate and the elevation. The locations and the elevations of the stations are presented in Figure~\ref{fig:stations}. To account for the differences in elevation, we construct radial basis functions also based on the elevation from the sea.

\subsection{Interpretation of the latent components}
\begin{figure}
  \centering
  \includegraphics[width=0.7\textwidth]{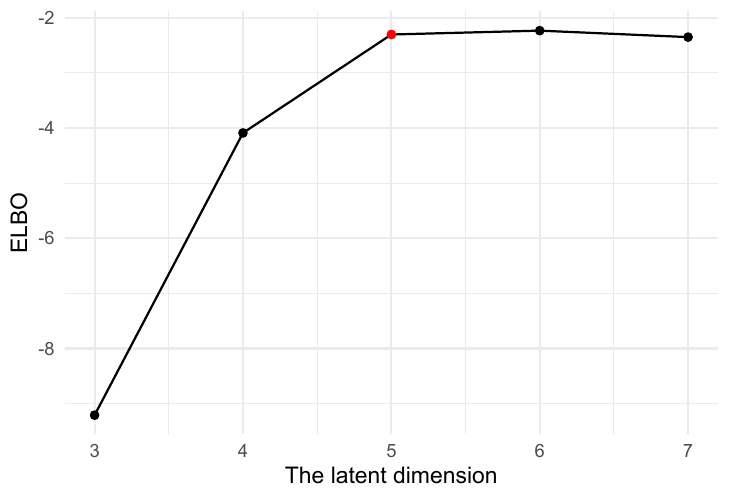}
  \caption{ELBOs for iVAEr models fitted with the latent dimensions $P=3, \dots, 7$. The selected latent dimension $P=5$ is marked as red.}
  \label{fig:ic_est_veneto}
\end{figure}

In this section, we use same iVAEr hyperparameters as in Section~\ref{sec:simulations}, but we add the radial basis functions based on the elevation. For elevation, we use resolution levels $E=(2, 9)$.
We begin with estimating the number of latent components by fitting iVAEr with latent dimensions $P = 3, \dots, 7$ and comparing the ELBOs. For each latent dimension, the method is run for 30 epochs. The ELBOs are presented against different latent dimensions in Figure~\ref{fig:ic_est_veneto}. The figure shows a clear knee point at $P=5$ after which the ELBO remains stable. The lowest uAIC is also obtained at $P=5$. Hence, we select the latent dimension $P=5$ for further analysis.

Next, iVAEr is fitted to the whole dataset by training the model for 60 epochs. To interpret the obtained latent components, we calculate the scaled mean absolute Shapley additive explanations (MASHAP) \cite{SipilaNonlinearSBSS} for fitted iVAEr's decoder. The scaled MASHAP are based on Shapley additive explanations (SHAP) \cite{kernelshap} and MASHAP values \cite{marcilio2020explanations}, but modified for obtaining population level feature importances for functions with vector valued output. The scaled MASHAP values can be interpreted as feature importance values for function's input variables, where higher value means that the input variable has more importance for the output variable. By averaging over the scaled MASHAP values for each input variable, the average scaled MASHAP values are obtained, which can be interpreted as population level importances for functions with multiple outputs.

\begin{figure}[ht]
    \centering
    \includegraphics[width=.75\linewidth]{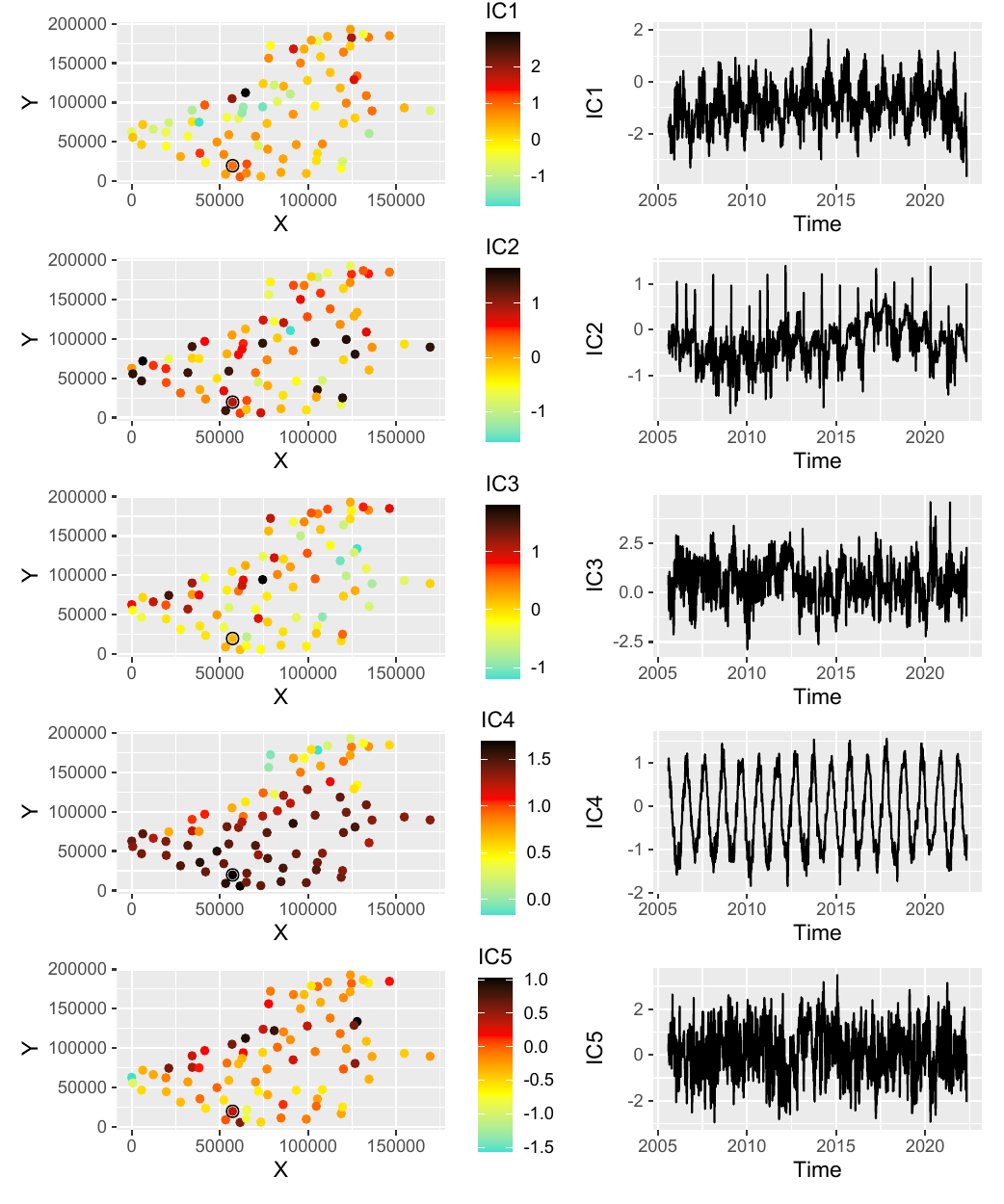}
  \caption{Spatial map of ICs (left column) for the first time point and a time series of ICs (right column) for the point circled in the spatial maps.} 
  \label{fig:ics_space_time}
\end{figure}

The scaled MASHAP values are calculated for decoder part by using 500 randomly selected observations as background data. The MASHAP values for the decoder part are presented in Table~\ref{tab:shap}. Spatial and temporal behaviours of the latent components are illustrated in Figure~\ref{fig:ics_space_time}. Based on the average scaled MASHAP values, the latent components IC4 and IC5 explain the most of the observed data.

\begin{table}
\caption{The scaled MASHAP and the average scaled MASHAP values calculated for iVAEr's decoder after the training process.}
\centering
\label{tab:shap}
\begin{tabular}{rrrrrr}
  \hline
 & IC1 & IC2 & IC3 & IC4 & IC5 \\
  \hline
ET\_0 & 0.125 & 0.026 & 0.029 & 0.721 & 0.099 \\
  T\_max & 0.037 & 0.036 & 0.090 & 0.722 & 0.115 \\
  T\_min & 0.148 & 0.029 & 0.095 & 0.659 & 0.069 \\
  H\_max & 0.170 & 0.026 & 0.484 & 0.037 & 0.284 \\
  H\_min & 0.129 & 0.045 & 0.280 & 0.160 & 0.386 \\
  log\_prec & 0.031 & 0.014 & 0.214 & 0.190 & 0.551 \\
  wind\_vel & 0.038 & 0.798 & 0.106 & 0.029 & 0.028 \\
   \hline
  Average & 0.097 & 0.139 & 0.185 & \textbf{0.360} & \textbf{0.219} \\ \hline
\end{tabular}
\end{table}

IC1 has small scaled MASHAP values for all variables, which indicates that IC1 is more of a residual component. It still has some seasonal variability based on the temporal behaviour.
IC2 explains only the wind. It has some seasonal variability and high peaks in time irregularly and does not show any clear spatial behaviour. The wind is not highly present in any other components meaning that wind might not share any common latent components with the other variables.
IC3 and IC5 together explain the most of the precipitation and humidity. IC3 has the smallest values in east while IC5 has the lowest ones in north-west. Both components seem to have higher values in the mountain area. The components do not have clear trend in time but differently from IC3, IC5 captures high frequency oscillations in time. 
IC4 captures the seasonal changes as well as elevation based spatial changes. When looking at temporal changes, it has low values during the winter and high values during the summer. Moreover, the low values occur in spatial locations where elevation is high and the high values occur in lowlands. In addition, the coastal area (on the right side of the map) seem to have slightly lower values than inland (the center and the left side of the map). Evapotranspiration, maximum temperature and minimum temperature have high scaled MASHAP values in IC4 and low values in other ICs, which indicates that IC4 mostly explains these variables. In addition, IC4 explains some of precipitation and minimum humidity.

\subsection{Spatio-temporal predictions}

\mika{In the next, we study if preprocessing the data with iVAEr improves spatio-temporal prediction accuracy. For comparison purposes we use STBSS \cite{MuehlmannDeIacoNordhausen2022} and STLCM approaches \cite{deiaco2003, deiaco2005}. In particular, we apply STBSS as preprocessing method and direct predicting using spatio-temporal kriging \cite{kyriakidis1999geostatistical}, as well as we use the STLCM to forecast the target variables using cokriging \cite{Wackernagel2003}.
%As comparative methods we use STBSS as preprocessing method, direct predicting using spatio-temporal kriging \cite{kyriakidis1999geostatistical} and linear coregionalization model \cite{deiaco2003, deiaco2005, deiaco2012, deiaco2013a, deiaco2013b, deiaco2019, deiaco2022} with spatio-temporal cokriging. 
We consider four different iVAEr approaches to obtain the latent components $\bs z$.} The approaches are composed of two different iVAEr's auxiliary function settings; one using three hidden layers with 128 units (iVAEaux3) and the other using one hidden layer with \mika{16} units (iVAEaux1). The first two of the approaches use iVAEaux3 and iVAEaux1 directly to predict the latent components to new spatio-temporal locations, and the other two approaches combine iVAEaux3 and iVAEaux1 with spatio-temporal kriging to obtain the predictions. When iVAEaux3 and iVAEaux1 are used directly, the trend function $\bs \mu(\bs s, t)$ provided by the auxiliary function, is treated directly as the predictions for the latent components. When iVAEaux3 and iVAEaux1 are combined with kriging, the latent components $\bs z(\bs s, t)$ are preprocessed further by subtracting the mean function $\bs \mu(\bs s, t)$ as $\bs z_{res}(\bs s, t) = \bs z(\bs s, t) - \bs \mu(\bs s, t)$. The residuals $\bs z_{res}$ are then predicted to new locations using spatio-temporal kriging, and finally, the trend is added back to the predictions as $\bs z(\bs s_{new}, t_{new}) = \bs z_{res}(\bs s_{new}, t_{new}) + \bs \mu(\bs s_{new}, t_{new})$. To obtain the predictions in the original observation space, the predicted latent components are back transformed using iVAEr's mixing function estimate.

For prediction purposes, the auxiliary variable should be formed in a way that the scope of auxiliary variables for the prediction locations is not far out of the scope of the auxiliary variables for the training locations. As our training data has no information of the future time points, the spatio-temporal trend function is not reliable for future time points in regular iVAEr setup. To make the scope same for both training and validation data, we utilize the seasonality of the data instead of using the absolute time points to form the temporal radial basis functions. To account for the seasonality, the temporal radial basis functions are formed using the week of the year (1-53) instead of using the weeks from the first observation (1-892). In addition, to allow the differences between different years, we add a one-hot encoded year factor, i.e. a 18-dimensional standard basis vector giving the year of the observation, to the auxiliary variable. With such setup, the auxiliary variables used for predictions are not out of scope of the variables used for training and hence, iVAE's auxiliary function is able to predict the trend to the future better than by using the regular setup with temporal radial basis functions based on the absolute time points. The auxiliary function of iVAEaux3 with three hidden layers has capability of learning finer details of the spatio-temporal field, but is also more prone to overfitting. The auxiliary function of iVAEaux1 on the other hand is less likely to overfit as it has much less parameters and might be for that reason better for trend estimation. In each iVAE setting, we use the spatial resolution levels $H = (2, 9)$, temporal (seasonal) resolution levels $G=(2, 9)$ and elevation based resolution levels $E=(2, 9)$.

\mika{The reference methods, direct spatio-temporal kriging, STBSS combined with kriging and STLCM with cokriging, expect the data without seasonal components.} Since the weather data has clear seasonality, we hence preprocess the data by subtracting the estimated seasonal component \mika{from each original variable}. The seasonal cyclicality is estimated by fitting the model 
\begin{align*}    
x_i(\bs s, t) = \beta_{0,i} + \beta_{1,i}\text{cos}(2 \pi t / 53) + \beta_{2,i}\text{sin}(2 \pi t / 53) + x_{res,i}(\bs s, t)
\end{align*}
for each variable $x_i$, $i=1,\dots,7$. \mika{In direct kriging approach, the residuals $x_{res,i}$ are predicted using spatio-temporal kriging. In STBSS based approach, STBSS is applied to the residuals to obtain the latent components. The latent components are then predicted using kriging, and the predictions are backtransformed to original data using STBSS. The parameters for STBSS are the best performing parameters based on the simulation studies, but scaled to match the spatial domain of Veneto dataset.
In the STLCM, we first have selected 3 latent components identified by joint diagonalization of the sample covariance matrices \cite{deiaco2012,deiaco2013a,deiaco2019,deiaco2022}. 
Then, supported by the evaluation of the non-separability index \cite{cappello2020, deiaco2019isotropy, deiaco2013positive} the product-sum covariance model has been chosen to describe the characteristics of the covariance functions estimated on the three retained components. Finally, the STLCM model has been used to compute spatio-temporal predictions of the seven analyzed variables.}\\
To estimate the seasonal components $\mu_i(\bs s, t) = \hat{\beta}_{0,i} + \hat{\beta}_{1,i}\text{cos}(2 \pi t / 53) + \hat{\beta}_{2,i}\text{sin}(2 \pi t / 53)$ for new spatial locations we use the modified version of the approach used in \cite{deiaco2012} for each variable $x_i$, $i = 1, \dots, 7$:
\begin{enumerate}
    \item Calculate means $\mu_i(\bs s_j)$ and standard deviations $\sigma_i(\bs s_j)$ of the seasonal components for the training stations $\bs s_j$, $j=1,\dots, n_s$. Standardize the seasonal components for each training station to have zero mean and unit variance as $\tilde\mu_i(\bs s_j, t) = \frac{\mu_i(\bs s_j, t) - \mu_i(\bs s_j)}{\sigma_i(\bs s_j)}$.
    \item Fit the periodic function $\tilde\mu_i(\bs s_j, t) = \tilde\beta_{1, i}\text{cos}(2 \pi t / 53) + \tilde\beta_{2, i}\text{sin}(2 \pi t / 53) + \tilde\mu_{res, i}(\bs s_j, t)$ for the standardized seasonal components to obtain the periodic component $\tilde\mu_i(t) = \hat{\tilde\beta}_{1, i}\text{cos}(2 \pi t / 53) + \hat{\tilde\beta}_{2, i}\text{sin}(2 \pi t / 53)$.
    \item Consider the means $\mu_i(\bs s_j)$ and standard deviations $\sigma_i(\bs s_j)$ as realizations of two spatial random fields. Use spatial kriging to estimate the mean and the variance to new locations $\bs s_{new}$.
    \item Construct the trend to location $\bs s_{new}$ as $\mu_i(\bs s_{new}, t)$ = $\tilde\mu_i(t)\sigma_i(\bs s_{new}) + \mu_i(\bs s_{new})$.
\end{enumerate}
The final predictions $x_i(\bs s_{new}, t_{new})$ are then obtained as $x_i(\bs s_{new}, t_{new}) = \linebreak x_{res,i}(\bs s_{new}, t_{new}) + \mu_i(\bs s_{new}, t_{new})$, where $x_{res,i}(\bs s_{new}, t_{new})$ is the predicted residual of the deseasonalized observation. In kriging step for both original variables and for latent components \mika{provided by iVAEr and STBSS, we fit either integrated product sum covariance model \cite{de2002nonseparable} or the product sum covariance model \cite{de2001space} based on the non-separability index.} The kriging predictions are calculated using 40 nearest points.

As measures of performance, we calculate mean squared error $MSE(\bs x_i, \hat{\bs x}_i) = \frac{1}{n}\sum_{j=1}^n(x_{i,j} - \hat{x}_{i, j})^2$ and mean absolute error $MAE(\bs x_i, \hat{\bs x}_i) = \frac{1}{n}\sum_{j=1}^n |x_{i,j} - \hat{x}_{i, j}|$, where $\bs x_i$ and $\hat {\bs x}_i$ has the true observations and predicted counterparts of $i$th observed variable, respectively. In addition, we calculate weighted average MSE (wMSE) and weighted average MAE (wMAE), which are calculated as \mika{$wMSE(\bs X, \hat{\bs X}) = \frac{1}{7}\sum_{i = 1}^7 \frac{MSE(\bs x_i, \hat{\bs x}_i)}{\sigma^2(\bs x_i)}$ and $wMAE(\bs X, \hat{\bs X}) = \frac{1}{7}\sum_{i = 1}^7 \frac{MAE(\bs x_i, \hat{\bs x}_i)}{\sigma(\bs x_i)}$}, where $\bs X$ is a $n \times 7$ matrix containing the true observation vectors as rows, $\hat{\bs X}$ has the predicted observations and \mika{$\sigma(x_i)$} is a standard deviation of $i$th observed variable calculated from the \mika{deseasonalized train} data. wMSE and wMAE account for the scale differences of the variables and give a single measure of performance regarding the whole data.

\begin{table}
\centering
\caption{Mean squared errors of iVAE with three hidden layers in auxiliary function (iVAEaux3), iVAEaux3 combined with kriging, iVAE with one hidden layer in auxiliary function (iVAEaux1), iVAEaux1 combined with kriging, regular univariate spatio-temporal kriging, \mika{STBSS combined with kriging and STLCM combined with cokriging}. The smallest errors are bolded for each variable.}
\label{tab:prediction_results_mse}
\bigskip
\resizebox{0.750\textwidth}{!}{
\subfloat[MSEs and wMSE for temporal part]{
\begin{tabular}{rrrrrrrr|r}
  \hline
 & ET\_0 & T\_max & T\_min & H\_max & H\_min & log\_prec & wind\_vel & wMSE\\
  \hline
  iVAEaux3 & \textbf{0.06} & \textbf{3.72} & 4.21 & 87.28 & 175.69 & 2.12 & 18.15 & 1.11 \\    
  iVAEaux3 + kriging & 0.07 & 4.47 & 4.29 & 85.00 & \textbf{165.08} & 2.15 & \textbf{16.20} & \textbf{1.10} \\
  iVAEaux1 & 0.07 & 4.98 & 5.24 & 117.92 & 173.91 & 2.58 & 22.83 & 1.38 \\ 
  iVAEaux1 + kriging & 0.11 & 7.90 & 5.21 & 115.73 & 235.22 & 2.02 & 20.59 & 1.46 \\ 
  Kriging & 0.12 & 6.28 & \textbf{3.49} & 89.37 & 358.14 & 1.97 & 19.38 & 1.41 \\
  \mika{STBSS + kriging} & 0.11 &  8.04 & 3.51 &  91.69 & 335.65 &   \textbf{1.95} & 16.70 & 1.38 \\
  \mika{STLCM + cokriging} & 0.08  &  5.55  &  3.84 & \textbf{79.66}  & 207.99  &  1.96  &  17.83  & 1.15   \\
   \hline
\end{tabular}}}

\resizebox{0.750\textwidth}{!}{
\subfloat[MSEs and wMSE for spatial part]{
\begin{tabular}{rrrrrrrr|r}
  \hline
 & ET\_0 & T\_max & T\_min & H\_max & H\_min & log\_prec & wind\_vel & wMSE \\
  \hline
  iVAEaux3 & 0.10 & 5.33 & 6.61 & 74.45 & 123.91 & 1.84 & 104.05 & 2.16 \\
  iVAEaux3 + kriging & 0.07 & 2.27 & 4.00 & 52.43 & 44.78 & 0.28 & 125.16 & 1.98 \\
  iVAEaux1 & 0.12 & 6.72 & 9.11 & 65.07 & 141.78 & 2.19 & 41.73 & 1.52 \\ 
  iVAEaux1 + kriging & \textbf{0.05} & \textbf{1.55} & 3.22 & \textbf{33.82} & \textbf{33.14} & 0.25 & \textbf{38.34} & \textbf{0.81} \\
  Kriging & 0.19 & 8.22 & 2.47 & 37.90 & 57.66 & 0.26 & 39.06 & 1.10 \\
  \mika{STBSS + kriging} & 0.19  &   8.30  &   2.49  &  36.52  &  59.76   &  0.30  &  39.00   &  1.11 \\
    \mika{STLCM + cokriging} & 0.17  &   8.20  &   \textbf{2.42}  &  37.76  &  56.74   &  \textbf{0.23}  &  39.02   & 1.08  \\
   \hline
\end{tabular}}}

\resizebox{0.750\textwidth}{!}{
\subfloat[MSEs and wMSE for spatio-temporal part]{
\begin{tabular}{rrrrrrrr|r}
  \hline
 & ET\_0 & T\_max & T\_min & H\_max & H\_min & log\_prec & wind\_vel & wMSE \\
  \hline
  iVAEaux3 & \textbf{0.07} & 3.65 & 6.64 & 249.94 & 258.02 & 2.37 & 106.01 & 2.97 \\ 
  iVAEaux3 + kriging & \textbf{0.07} & \textbf{3.57} & 6.98 & 249.96 & 251.24 & 2.42 & 123.50 & 3.19 \\
  iVAEaux1 & \textbf{0.07} & 5.47 & 7.24 & 208.56 & 206.50 & 2.53 & 35.80 & 1.99 \\ 
  iVAEaux1 + kriging & 0.09 & 6.57 & 6.62 & 187.59 & \textbf{177.25} & 2.25 & 33.83 & \textbf{1.86} \\
  Kriging & 0.28 & 14.40 & 6.57 & 207.04 & 409.58 & 2.13 & 58.01 & 2.79 \\
  \mika{STBSS + kriging} & 0.20  &  12.98  &   \textbf{5.88}  & 199.71  & 334.76   &  2.02  &  \textbf{32.43} & 2.24 \\
  \mika{STLCM + cokriging} & 0.18  &  10.41  &   6.06  & \textbf{184.22}  & 251.47   & \textbf{1.94}  &  33.82  & 2.05 \\
   \hline
\end{tabular}}}
\end{table}

\begin{table}
    \centering  
\caption{Mean absolute errors of iVAE with three hidden layers in auxiliary function (iVAEaux3), iVAEaux3 combined with kriging, iVAE with one hidden layer in auxiliary function (iVAEaux1), iVAEaux1 combined with kriging, regular univariate spatio-temporal kriging, \mika{STBSS combined with kriging and STLCM combined with cokriging}. The smallest errors are bolded for each variable.}
\label{tab:prediction_results_mae}
\bigskip
\resizebox{0.75\textwidth}{!}{
\subfloat[MAEs and wMAE for temporal part]{
\begin{tabular}{rrrrrrrr|r}
  \hline
 & ET\_0 & T\_max & T\_min & H\_max & H\_min & log\_prec & wind\_vel & wMAE \\
  \hline
  iVAEaux3 & \textbf{0.19} & \textbf{1.63} & 1.65 & 6.05 & 10.73 & 1.22 & 2.77 & \textbf{0.76} \\       
  iVAEaux3 + kriging & 0.20 & 1.75 & 1.68 & \textbf{5.99} & \textbf{10.41} & 1.25 & \textbf{2.55} & \textbf{0.76} \\ 
  iVAEaux1 & 0.21 & 1.84 & 1.78 & 7.04 & 10.77 & 1.45 & 3.52 & 0.86 \\ 
  iVAEaux1 + kriging & 0.25 & 2.21 & 1.85 & 7.04 & 12.34 & \textbf{1.17} & 3.06 & 0.88 \\
  Kriging & 0.27 & 2.06 & 1.54 & 6.11 & 15.67 & 1.26 & 3.10 & 0.88 \\
  \mika{STBSS + kriging} & 0.25  &   2.34  &   \textbf{1.52}  &   6.22  &  14.98   &  1.23   &  2.71   &  0.87 \\ 
    \mika{STLCM + cokriging} & 0.24  &   1.94 &   1.60  &   6.09  &  11.66  &  1.26   &  2.86   &  0.81 \\ 
   \hline
\end{tabular}}}

\resizebox{0.75\textwidth}{!}{
\subfloat[MAEs and wMAE for spatial part]{
\begin{tabular}{rrrrrrrr|r}
  \hline
 & ET\_0 & T\_max & T\_min & H\_max & H\_min & log\_prec & wind\_vel & wMAE \\
  \hline
  iVAEaux3 & 0.24 & 1.80 & 2.04 & 5.51 & 8.35 & 1.10 & 7.63 & 0.96 \\
  iVAEaux3 + kriging & 0.20 & 1.14 & 1.49 & 4.83 & 4.49 & 0.37 & 8.36 & 0.77 \\
  iVAEaux1 & 0.25 & 2.05 & 2.39 & 5.06 & 9.33 & 1.25 & 5.19 & 0.91 \\ 
  iVAEaux1 + kriging & \textbf{0.16} & \textbf{0.96} & 1.28 & \textbf{3.69} & \textbf{4.18} & 0.33 & \textbf{4.93} & \textbf{0.56} \\ 
  Kriging & 0.27 & 1.98 & \textbf{1.17} & 4.19 & 5.78 & 0.36 & 5.16 & 0.70 \\
  \mika{STBSS + kriging} & 0.28   &  1.99   &  1.18    & 4.15   &  5.86    & 0.41  &   5.16   &  0.70 \\
  \mika{STLCM + cokriging} & 0.26  &   1.89 &   \textbf{1.17}  &   4.17  &  5.79  &  \textbf{0.30}   &  5.11   &  0.68 \\ 
   \hline
\end{tabular}}}

\resizebox{0.75\textwidth}{!}{
\subfloat[MAEs and wMAE for spatio-temporal part]{
\begin{tabular}{rrrrrrrr|r}
  \hline
 & ET\_0 & T\_max & T\_min & H\_max & H\_min & log\_prec & wind\_vel & wMAE \\
  \hline
  iVAEaux3 & \textbf{0.20} & \textbf{1.59} & 2.19 & 10.92 & 12.58 & 1.31 & 7.59 & 1.14 \\ 
  iVAEaux3 + kriging & \textbf{0.20} & \textbf{1.59} & 2.26 & 10.91 & 12.44 & 1.34 & 8.27 & 1.17 \\
  iVAEaux1 & 0.22 & 1.89 & 2.05 & 9.70 & 11.68 & 1.43 & \textbf{4.21} & 0.99 \\ 
  iVAEaux1 + kriging & 0.23 & 2.10 & 2.05 & \textbf{9.36} & \textbf{10.96} & 1.28 & 4.26 & \textbf{0.97} \\
  Kriging & 0.36 & 3.04 & 2.15 & 9.71 & 16.70 & 1.30 & 6.12 & 1.24 \\
  \mika{STBSS + kriging}  & 0.32   &  2.92  &   \textbf{1.99}  &   9.85  &  15.11   &  1.29  &   4.50  &  1.12 \\
    \mika{STLCM + cokriging} & 0.32  &   2.43  & 2.07  &   9.78  &  12.67   &  \textbf{1.26}   &  4.67   & 1.07 \\ 
   \hline
\end{tabular}}}
\end{table}

The validation prediction errors are collected in Tables~\ref{tab:prediction_results_mse} and \ref{tab:prediction_results_mae}, where the cases (a), (b) and (c) has the prediction errors for temporal part, spatial part and spatio-temporal part, respectively. When the observed variables are predicted to future (cases (a) in Tables~\ref{tab:prediction_results_mse} and \ref{tab:prediction_results_mae}) considering only the spatial locations present in the training data, the best overall results are obtained using iVAEaux3 or iVAEaux3 + kriging. \mika{They have slightly lower wMSE and wMAE than the second best method, STLCM + cokriging, and clearly outperform the other methods. When the prediction errors of the individual variables are inspected, it is evident that iVAEaux3 and iVAEaux3 + kriging provide the lowest MSE values for all but minimum temperature, maximum humidity and log precipitation. The lowest MSE for these variables are obtained by kriging, STLCM + cokriging and STBSS + kriging, respectively. The lowest MAE for all variables but minimum temperature and log precipitation are also obtained by iVAEaux3 and iVAEaux3 + kriging. The lowest MAE for minimum temperature and log precipitation are obtained by STBSS + kriging and iVAEaux1 + kriging, respectively.}

When the observed variables are predicted to new spatial locations (cases (b) in Tables~\ref{tab:prediction_results_mse} and \ref{tab:prediction_results_mae}), but not to the future (i.e. using all time points from the training data), the best overall results are obtained by iVAEaux1 + kriging as it has significantly lower wMSE and wMAE than any of the competing methods. \mika{iVAEaux1 + kriging has the lowest MAE and MSE for evapotranspiration, maximum temperature, maximum humidity and minimum humidity and wind velocity. The lowest MSE and MAE for minimum temperature and log precipitation are obtained by STLCM + cokriging.}

When the observed variables are predicted to future and to new spatial locations (cases (c) in Tables~\ref{tab:prediction_results_mse} and \ref{tab:prediction_results_mae}), \mika{iVAEaux1 + kriging is the best performing method, followed by iVAEaux1 and STLCM + cokriging. The lowest wMSE and wMAE are both obtained by iVAEaux1 + kriging. However, when inspecting MSE and MAE values for individual variables, the results are spread more between the methods. All iVAE based methods have the lowest MSE and MAE values for evapotranspiration. For maximum temperature the differences in the errors are high. The lowest errors are obtained by iVAEaux3 based methods, followed by iVAEaux1 based methods, while for minimum temperature, STBSS + kriging has the lowest value, but the differences between the errors are low. The lowest MSE for maximum humidity is obtained by STLCM + cokriging and the lowest MAE by iVAEaux1 + kriging. MSE and MAE for minimum humidity are clearly lowest for iVAEaux1 + kriging. For log precipitation and wind velocity, the differences between the best performing methods small. The lowest errors for log precipitation are obtained by STLCM + cokriging, and for wind velocity the lowest MSE is obtained by STBSS + kriging and the lowest MAE by iVAEaux1.} 

In conclusion, it is evident that iVAEr, estimating the latent components and spatio-temporal trend (and variance) function simultaneously, clearly improves the prediction accuracy compared to \mika{competing methods, especially in spatial data case}. For temporal prediction, the best accuracy is obtained by iVAEaux3 using the auxiliary function with three hidden layers. This hints that for temporal prediction, it is beneficial to capture finer details of the spatio-temporal structure. For temporal part, kriging did not improve significantly the prediction accuracy meaning that it is sufficient to use solely iVAEr in this situation. For spatial prediction, in the other hand, iVAEaux1, \mika{using one hidden layer with only 16 neurons} in auxiliary function, provides better results than iVAEaux3 hinting that it is more beneficial to capture only larger scale spatio-temporal structure and predict the residuals further using spatio-temporal kriging. \mika{In addition to performance gains of iVAEr based methods, they benefit from more simple modelling process as there is no need for fitting and predicting the spatio-temporal trend function separately as this is done simultaneously by the algorithm. Moreover, by providing independent components, the prediction can be done efficiently using univariate prediction methods such as kriging, which is computationally more manageable as compared to multivariate prediction methods such as cokriging.} \mika{Although even in this last approach, the multivariate estimation is often simplified by modelling uncorrelated components, identified through the joint diagonalization of the covariance matrices.}

\section{Conclusion and discussion}
\label{sec:conclusion}

In this paper, iVAE was extended to the nonstationary spatio-temporal setting, and three approaches, coordinate based, segmentation based and radial basis function based, were introduced for constructing the spatio-temporal auxiliary data. In addition, two latent dimension estimation methods were proposed. The introduced spatio-temporal iVAE and latent dimension estimation methods were studied using vast simulation studies and illustrated in meteorological application where also a novel iVAEr preprocessing procedure for accounting nonstationarity in spatio-temporal modelling was introduced. 

Based on the simulations, iVAEr, iVAEs1 and iVAEs2 were the best performing methods in nonlinear STBSS settings with nonstarionary spatio-temporal variance. The methods outperformed iVAEc, iVAEs3 and STBSS methods in all settings. FICA was still the best performing method under the linear mixing in settings without spatio-temporal trend and highly nonlinear variance. However, under nonlinear mixing or when the trend was present, iVAE methods outperformed FICA. Based on the fact that iVAEr provides for the latent components smooth trend and variance function estimates, which can be useful for further analysis, we consider iVAEr the best method for nonstationary STBSS.

In meteorological application, we utilized the introduced latent dimension estimation methods and iVAEr to find the underlying latent components. We interpreted the components using scaled MASHAP values and by inspecting the spatial and temporal behaviours of the components. Original seven variables were compressed into five latent components; one explained the seasonal and spatial variability in temperature and evapotranspiration, one explained the wind velocity, two explained together the most of the precipitation and humidity and the last one explained the remaining residuals of the data. For spatio-temporal prediction purposes, we utilized iVAEr preprocessing by estimating the latent components and their nonstationary spatio-temporal trend functions, which were used to account nonstationarity in the modelling of the latent components. When using iVAEr preprocessing, the prediction accuracy was improved as compared to predicting the original variables directly \mika{(through kriging or cokriging)} \mika{or to using STBSS as preprocessing method}.

\mika{Based on the results and theoretical properties of the developed spatio-temporal iVAE methods, it is evident that the methods have multiple advantages over the previously proposed STBSS methods. The methods are capable of estimating nonlinear mixing and unmixing functions in nonstationary data settings, whereas the previously proposed methods are designed only for linear mixing and stationary data settings. The developed iVAE methods can estimate injective mixing function, while previous linear methods assume that the latent and the observed dimensions are equal. In addition, iVAEr is highly useful for spatio-temporal modelling and prediction as it can estimate nonstationary spatio-temporal trend and variance functions. By removing the nonstationary trend from the components, stationary prediction methods, such as kriging, can provide better predictions as seen in the meteorological application.}

\mika{Although in this paper we performed a small simulation study regarding sensitivity of iVAEr against different radial basis function settings, more in depth study of sensitivity against different hyperparameters or source density mismatch is needed and will be done in future.} The spatio-temporal iVAE methods developed in this paper rely on nonstationary variance for identifiability. As discussed in Section~\ref{sec:ilsa}, there are other possibilities for introducing nonstationarity in spatio-temporal settings. The methods for other scenarios, such as nonstationary autocorrelation, will be developed in the future. Also, nonlinear STBSS methods for stationary data are left for future work. Nonlinear SBSS and STBSS methods have so far been studied and developed mainly for Gaussian data. In future, different source densities will be considered and methods that are resistant to outliers will be developed in a similar manner as was done in linear BSS framework in \cite{robustSBSS}. In future, we will also focus on developing nonlinear BSS methods for graph data, where one cannot measure the distance between the observations as in temporal, spatial or spatio-temporal framework, but only the relations between the observations are known.

%\mika{Although we focused in this paper only on meteorological application, the spatio-temporal iVAE methods are applicable for any multivariate spatio-temporal data. For example, in the field of neuroimaging, magnetic resonance spectroscopic imaging (MRSI) can be used to measure multiple chemicals across the brain over some time period making the application multivariate spatio-temporal. In addition, many traditional neuroimaging applications such as electroencephalography (EEG) or magnetoencephalography (MEG) can be considered as univariate spatio-temporal data. If multiple similar clinical trials were performed to the same patient, the data can be seen as multivariate spatio-temporal data. In such applications, spatio-temporal iVAE could reveal underlying spatio-temporal patterns of brain function, which cannot be captured directly. To study versatility of the proposed iVAE methods further, various real data examples, such as neuroimaging or epidemiological patterns, should be considered in future.}

\mika{Although this paper focuses exclusively on a meteorological application, the spatio-temporal iVAE methods are applicable to any multivariate spatio-temporal data. For instance, in neuroimaging, magnetic resonance spectroscopic imaging (MRSI) measures multiple chemicals across the brain over a period of time, making it a multivariate spatio-temporal application. Similarly, many traditional neuroimaging techniques such as electroencephalography (EEG) or magnetoencephalography (MEG) can be viewed as univariate spatio-temporal data. If multiple similar clinical trials are performed on the same patient, the resulting data can be treated as multivariate spatio-temporal.

In such scenarios, spatio-temporal iVAE methods could uncover latent spatio-temporal patterns in brain function that are not directly observable. To further explore the versatility of the proposed iVAE methods, future work should include diverse real-world data examples, such as neuroimaging or epidemiological pattern studies.}

\section*{Acknowledgments}
We acknowledge the support from Vilho, Yrjö and Kalle Väisälä foundation for MS, the support from the Research Council of Finland (453691) to ST, the support from the Research Council of Finland (363261) to KN and the support from the HiTEc COST Action (CA21163) to KN and ST.

%Bibliography
\bibliographystyle{elsarticle-num} 
\bibliography{references}  

\appendix

\mika{
\section{Computational complexity analysis}
\label{sec:computational_complexity}
The computational complexities of the proposed algorithms are composed of two parts, forming the auxiliary data and training iVAE. We use Big O notation to represent the worst case time and space complexities, where $O(n)$ denotes linear growth in computation time or memory usage with respect to the input size $n$. First, let us address iVAE's computational complexity which is very similar to any feed forward neural network such as regular VAE. Using the Big O notation, the computational time complexity for training the model is $O(n \times n_w \times n_e)$, where $n$ is the sample size, $n_w$ is the number of weights in the model and $n_e$ is the number of epochs. When the sample size $n$ grows, less epochs are typically needed for training, which makes the model well scaleable in terms of sample size. The memory usage, i.e. space complexity, is $O(n_w)$ for storing the weights of the model. The number of weights $n_w$ can be broken down to number of weights $n_{w1}$ in encoder-decoder part and the number of weights $n_{w2}$ in auxiliary function. $n_{w1}$ and $n_{w2}$ are heavily dependent on the number and size of the hidden layers. Typically fairly small neural networks (e.g. 3 layers with 128 units) in encoder, decoder and auxiliary functions are sufficient. In addition, $n_{w1}$ depends linearly on the dimension of the input data and latent dimension whereas $n_{w2}$ depends linearly on the dimension of auxiliary data. Also, if the input dimension is very large (e.g. more than 100), a larger encoder-decoder network might be needed. Hence, the time complexity grows more when input dimension, latent dimension or auxiliary data dimension grow. 

In iVAEc, the time and space complexities are the lowest as the coordinates are only scaled to form the auxiliary variables, meaning that using Big O notation, both time and space complexities are $O(n)$ for forming and storing the auxiliary data. The auxiliary data is only two dimensional, which makes time complexity slightly lower compared to other algorithms. In iVAEs1-iVAEs3, the time complexity of forming the auxiliary data is $O(m_{\mathcal{S}_1} \times m_{\mathcal{S}_2} \times m_{\mathcal{T}} \times n)$, where $m_{\mathcal{S}_1}, m_{\mathcal{S}_2}$ and $m_{\mathcal{T}}$ are the number of segments along each dimension. However, for iVAEs1, where all dimensions are considered jointly, and hence the dimension of auxiliary data can be very large, the space complexity is $O(m_{\mathcal{S}_1} \times m_{\mathcal{S}_2} \times m_{\mathcal{T}} \times n)$. For iVAEs2, the space complexity is $O((m_{\mathcal{S}_1} \times m_{\mathcal{S}_2} + m_{\mathcal{T}}) \times n)$ and for iVAEs3, it is $O((m_{\mathcal{S}_1} + m_{\mathcal{S}_2} + m_{\mathcal{T}}) \times n)$. In terms of computation time, iVAEs3 is the most efficient of segmentation based algorithms as the auxiliary dimension is the lowest. iVAEs2 is also efficient if the number of spatial segments is not very high. In iVAEr, the time and space complexities for forming and storing auxiliary data are $O((K_\mathcal{S} + K_\mathcal{T}) \times n$, where $K_\mathcal{S}$ and $K_\mathcal{T}$ are numbers of spatial and temporal node points, respectively. In all above iVAE variants, the space complexity can be reduced further by constructing auxiliary variables batch-wise during training process. In conclusion, the algorithms are well scalable in terms of sample size $n$ and relatively well scalable in terms of dimensions of input data, latent data and auxiliary data (linear time complexity). However, if the auxiliary data are not formed batch-wise, memory consumption may grow large if dimension of auxiliary variable is very large. Since the algorithm is essentially composed of three feed forward neural networks, encoder, decoder and auxiliary function, standard parallelization methods such as data parallelism, which distributes the data batch-wise across multiple computation units, can be applied to further reduce the overall computation time.}

\section{Additional simulation results}
\label{sec:appendix}

\begin{figure*}
  \centering
  \includegraphics[width=1\textwidth]{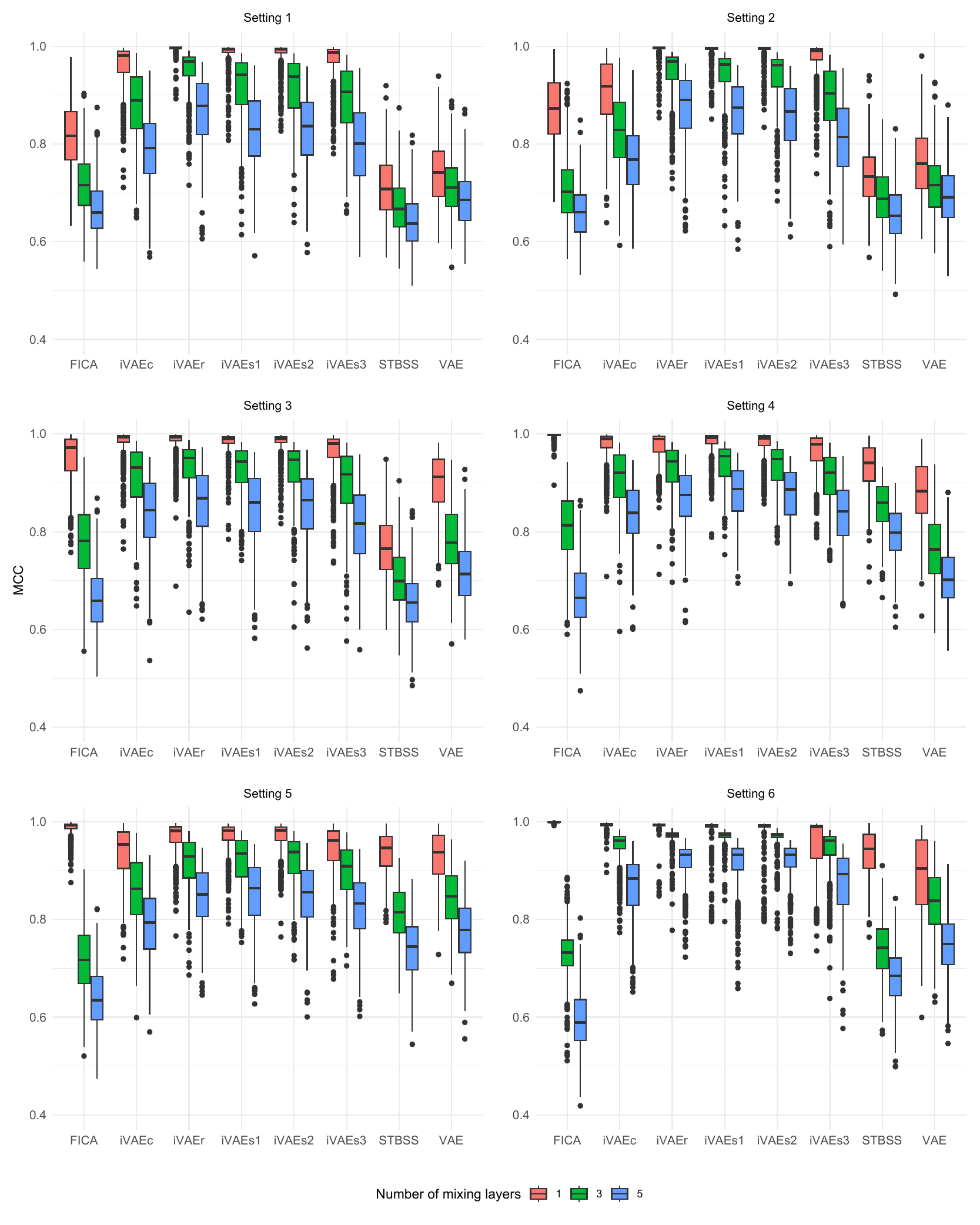}
  \caption{Mean correlation coefficients of 500 trials for Settings~1-6 for sample size with the number of spatial locations $n_s = 50$ and the number of temporal observations $n_t = 300$.}
  \label{fig:sim_all_s50t300}
\end{figure*}

\begin{figure*}
  \centering
  \includegraphics[width=1\textwidth]{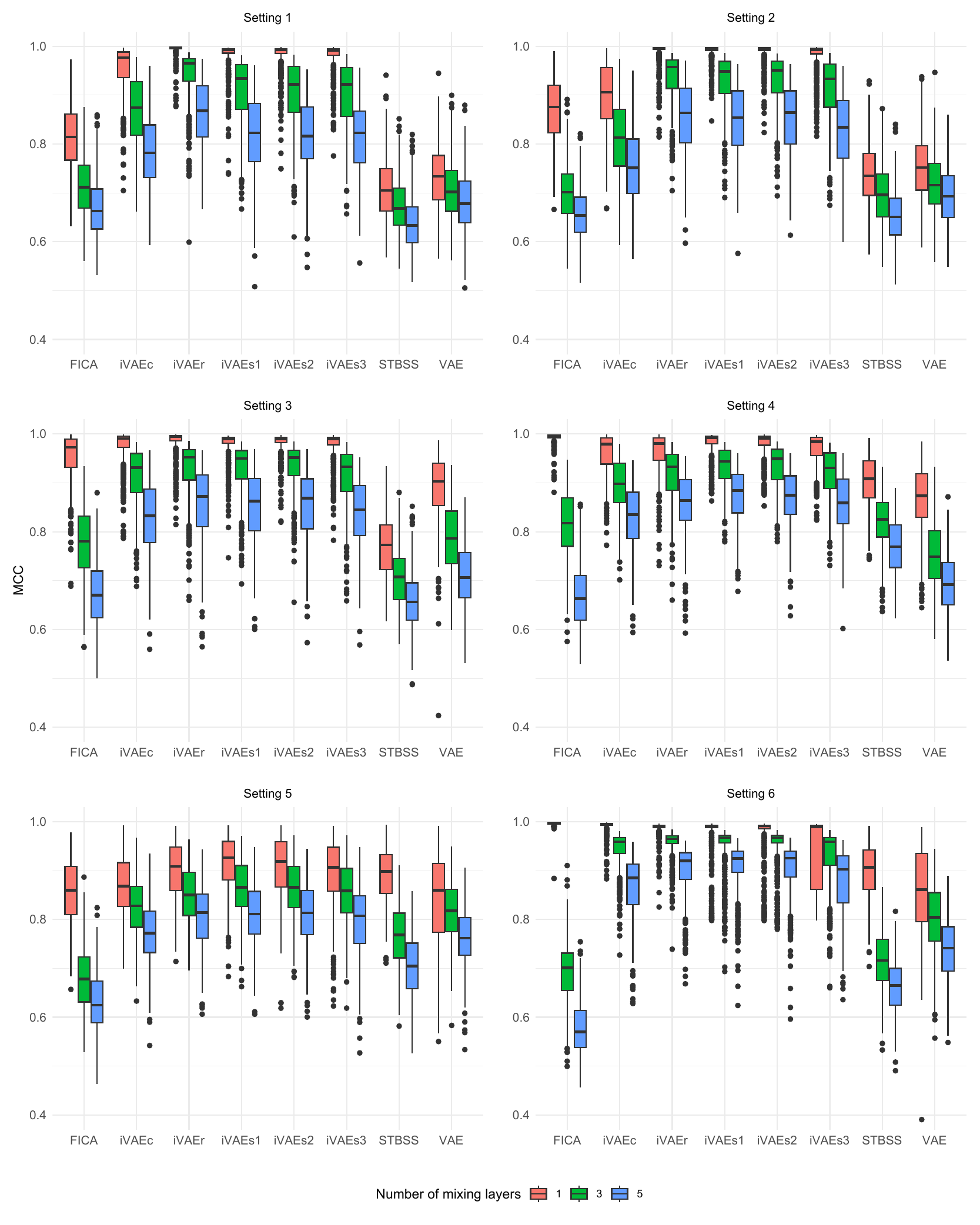}
  \caption{Mean correlation coefficients of 500 trials for settings~1-6 for sample size with the number of spatial locations $n_s = 150$ and the number of temporal observations $n_t = 75$.}
  \label{fig:sim_all_s150t75}
\end{figure*}

\begin{figure*}
  \centering
  \includegraphics[width=1\textwidth]{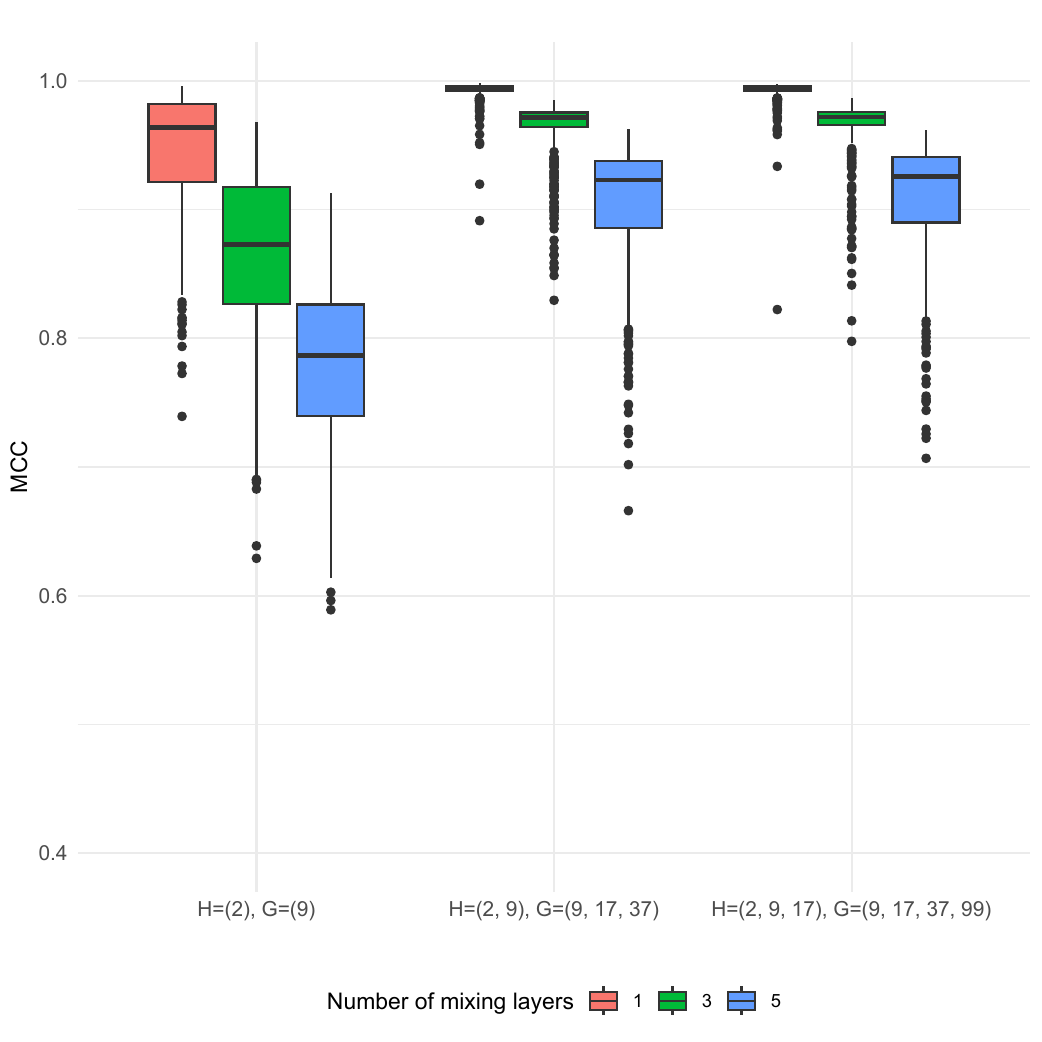}
  \caption{\mika{Mean correlation coefficients for different radial basis function parameter settings of iVAEr. The boxplots present 500 trials for Setting~6 with the number of spatial locations $n_s = 150$ and the number of temporal observations $n_t = 300$.}}
  \label{fig:radial_param_sensitivity}
\end{figure*}

\end{document}